%% file: ms.tex
\documentclass[11pt,preprint]{aastex}
\received{}
\accepted{}

\slugcomment{Accepted by ApJS}
\lefthead{Wilkes et al.}
\righthead{The SWIRE/\chandra\ Survey}

\usepackage{times,mathptm,graphicx,natbib}
\def\rl {R$_L$}
\def\q24 {q$_{24}$}
\def\sqdeg {deg$^2$}
\def\deg {$^{\rm o}$}

\def\etal   {{\rm et~al.}}
\def\alpe {$\alpha_E$}
\def\nh {${\rm N_H}$}
\def\lax    {${_<\atop^{\sim}}$}
\def\gax    {${_>\atop^{\sim}}$}
\def\chandra {{\it Chandra}}

\def\ang    {\rm \AA\/}
\def\fcgs {{erg cm$^{-2}$ s$^{-1}$}}

\begin{document}

\title{The SWIRE/\chandra\ Survey: The X-ray Sources.}
\author{Belinda J. Wilkes$^1$, Roy Kilgard$^2$, Dong-Woo Kim$^1$, Minsun
Kim$^3$, Mari Polletta$^4$, Carol Lonsdale$^5$, Harding E. Smith$^6$,
Jason Surace$^7$, Frazer N. Owen$^8$, A. Franceschini$^9$, Brian Siana$^{10}$
\& David Shupe$^{10}$
}
\affil{1: Harvard-Smithsonian Center for Astrophysics, Cambridge, MA 02138}
\affil{2: Department of Astronomy, Wesleyan University, CT}
\affil{3: Korea Astronomy and Space Science Institute (KASI), 61-1 ,
Hwaam-dong, Yuseong-gu, Deajeon 305-348 Korea}
\affil{4: INAF-ISAF  Milano, via E. Bassini 15, Milano 20133, Italy}
\affil{5: The National Radio Astronomy Observatory$^1$, Charlottesville, VI}
\affil{6: Center for Astrophysics and Space Science, University of
California, San Diego, La Jolla, CA 92093-0424}
\affil{7: {\it Spitzer} Science Center, California Institute of
Technology, 100-22, Pasadena, CA 91125}
\affil{8: National Radio Astronomy Observatory\footnote{The National Radio Astronomy
Observatory is a facility of the National Science Foundation operated
under cooperative agreement by Associated Universities Inc.}, P.O. Box, 0, Socorro, NM
87801 USA}
\affil{9: Dipartimento Di Astronomia, Universita di Padova, Vicolo
Osservatorio 5, I-35122 Padua, Italy}
\affil{10: California Institute of Technology, MC 105-24, 1200 East
California Boulevard, Pasadena, CA}

\begin{abstract}
We report a moderate-depth (70
ksec), contiguous 0.7  \sqdeg , \chandra\ survey, in the Lockman Hole Field
of the {\it Spitzer}/SWIRE Legacy Survey coincident with a
completed, ultra-deep VLA survey with deep optical and near-infrared (NIR) imaging
in-hand.  The primary motivation is to
distinguish starburst galaxies and Active Galactic Nuclei, including
the significant, highly obscured (log \nh $>$23) subset. \chandra\ has detected
775 X-ray sources to a limiting 
broad band (0.3-8 keV) flux $\sim 4 \times 10^{-16}$
\fcgs . We present the X-ray catalog, fluxes,
hardness ratios and multi-wavelength fluxes.
The log N vs. log S agrees with those of previous surveys
covering similar flux ranges. 
The Chandra and {\it Spitzer} flux limits are well matched: 771 (99\%) of the
X-ray sources have IR (infared) or optical counterparts,
and 333 have MIPS 24 $\mu$m detections.
There are 4 optical-only X-ray sources and 4 with no visible optical/IR
counterpart.
The very deep ($\sim$2.7 $\mu$Jy rms) VLA data yields
251 ($> 4 \sigma$) radio counterparts, 44\% of the X-ray sources in the field.
We confirm that the tendency for lower X-ray flux sources to be harder 
is primarily due to absorption.
As expected, there is no correlation between observed IR and X-ray
flux. Optically bright, Type 1 and red AGN lie in distinct
regions of the IR vs X-ray flux plots, demonstrating the wide range of
SEDs in this sample and providing the potential for classification/source
selection. Many optically-bright sources, which lie outside the AGN
region in the optical vs X-ray plots ($\rm f_r / f_x > 10$), lie inside the region
predicted for red AGN in IR vs X-ray plots consistent with
the presence of an active nucleus.
More than 40\% of the X-ray sources in the VLA
field are radio-loud using the classical definition, \rl . 
The majority of these
are red and relatively faint in the optical so that the use of \rl~to 
select those AGN with the
strongest radio emission becomes questionable. Using 
the 24 $\mu$m to radio flux ratio (\q24 ) 
instead results in 13 of the 147 AGN with sufficient data being
classified as radio-loud,
in good agreement with the $\sim$10\% expected for broad-lined AGN
based on optical surveys. 
We conclude that \q24 ~is a more reliable indicator of radio-loudness.
Use of \rl ~should be confined to optically-selected, Type 1 AGN.
\end{abstract}


\section{Introduction}
SWIRE, the largest {\it Spitzer} Legacy program, is tracing the evolution of dusty,
star-forming galaxies, evolved stellar populations and AGN, as a function
of environment from z \gax 2.5 to the present epoch.  SWIRE covers 6
fields with a total area of 49 \sqdeg\ in all seven {\it Spitzer} bands,
selected from the entire IRAS/DIRBE sky as those areas
with the lowest 100$\mu m$ surface brightness (which scales with $N_H$,
Schlegel \etal\ 1998); the Lockman Hole is one of the two best regions,
having several contiguous \sqdeg\ of low $N_H$($\sim 7 \times
10^{19}cm^{-2}$) sky.  SWIRE's power comes from
its large surface area, its depth,
which probes the Universe out to redshifts z \gax 2.5 and
over time intervals $> 10 Gyr$, and its sensitivity to both evolved stellar
populations (IRAC: 3--8$\mu m$) and dusty objects (starbursts \& obscured
AGN; MIPS: 24,70,160$\mu m$). Thus their
distribution relative to environment, and evolution with respect
to time and to the development of structure, can be studied together. 
SWIRE is one of several, on-going, medium-depth surveys which 
fill the gap between the deep and shallow surveys,
dramatically illustrating the need for a multi-layered approach to
efficiently fill the L-z plane in several wave-bands.

The SWIRE prime science goal is to 
study the structure, evolution and environments of AGN, starbursts and
spheroidal galaxies out to z\gax 2.5.
Key to this extensive multi-wavelength campaign is
an X-ray survey deep enough to distinguish starbursts and AGN,
including the significant numbers which are
highly obscured \nh\gax 10$^{23}$ cm$^{-2}$.
We carried out a \chandra\  X-ray survey  in
the best (in terms of Galactic extinction and
absence of nearby bright sources, including no bright radio sources) 
extragalactic $\sim$1 \sqdeg\
field within SWIRE. The X-ray observations cover 0.7 \sqdeg\
contiguously. The broad band (0.3-8.0 keV) flux limit is
$4 \times 10^{-16}$ \fcgs\ at the field centers,
sufficient to detect all SWIRE IR AGN except for those with high absorption
at low redshift (z$\leq$0.8, log N$_H \geq$ 24) while IR galaxies are
not detected. Thus we are able to distinguish all but
the most highly obscured AGN from amongst the IR sources by their X-ray
emission.

The standard AGN model includes a super-massive black hole
surrounded by an accretion disk in the center of a galaxy.
The central regions of the AGN produce strong, hard X-ray emission
(and relativistic jets in radio-loud sources) along with thermal
emission from an accretion disk. Gas in the vicinity is heated by the
nuclear emission and produces the broad and narrow emission lines
characteristic of Type 1 and Type 2 AGN respectively. 
The viewing orientation of an AGN
effects its classification, as demonstrated by the
detection of polarized broad lines in NGC1068 (Antonucci \& Miller 1985).
This result led to a general acceptance that some fraction of Type 2 AGN are
edge-on Type 1s. Such unification schemes require highly opaque dust
surrounding the accretion disk. In addition to
obscuring our view to the nucleus, this dust is heated by the nuclear source
producing strong IR emission. Thus the observed Spectral Energy Distribution
(SED) of an AGN is
dependent on viewing angle and the traditional optical/ultraviolet
surveys are incomplete to obscured sources ({\it e.g.} Polletta \etal\ 2006).
It is now clear that there is a significant 
obscured AGN population which has been missed from earlier surveys.
The powerful combination of radio, X-ray and IR observations available
for the current surveys facilitates a different and potentially
unbiased view of the AGN in the
field and so probes the full AGN population, including
obscured sources. The variety of
new techniques has resulted in new types of
AGN being found, including: red AGN (Cutri \etal\ 2002),
X-ray bright optically-normal galaxies (XBONGS, Comastri et al. 2002),
Type 2 quasars (Norman et al. 2002), X-ray
detected Extremely Red Objects (V/EROs, Alexander \etal\ 2002, Brusa \etal\ 2005).

Despite the several multi-wavelength surveys in progress
during this era of Great Observatories, 
a full view of the AGN population remains elusive.
Shallow surveys, which are dominated by Type 1 AGN, find
largely distinct subsets of the population depending on waveband (Hickox et
al. 2009). While overlap is significant in deeper surveys, selection in
any single waveband does not find all the AGN (Barmby et al 2006, Polletta et al.
2006, Donley et al. 2008, Park et al. 2008).
Samples of obscured AGN remain relatively small and biases are still
in the process of being understood, so that the nature 
of the new AGN and their significance to the 
population as a whole remain undetermined. Compton thick AGN are 
hard to find because their X-ray flux is obscured to energies \gax 10 keV.
Estimates of the fraction of the general population which are Compton thick
vary. X-ray selected AGN include $\sim 30$\% (Polletta \etal\ 2006, Treister et al.
2004) Compton-thick sources. This number doubles to $\sim 66\%$
when IR-selected AGN are included (Polletta et al. 2006, Treister et
al. 2009). 
Estimates generated by stacking X-rays from IR-selected AGN also suggest
a factor of two increase in IR-selected compared with X-ray
selected Compton thick AGN (Fiore et al. 2008, Daddi et al.
2007), although the absolute numbers differ by factors of $\sim$100 
depending on selection techniques and assumptions.
The latest estimate from modeling the Cosmic X-ray
Background (CXRB) postulates an obscured population larger than the
unobscured by a factor of 8 at low luminosities, half of which are Compton
thick. This decreases to a factor of 2 at high
luminosities (Gilli, Comastri \& Hasinger 2007).
Alternatively, many X-ray-selected, low-luminosity AGN may
be intrinsically X-ray hard rather than obscured, possibly reducing
the fraction of obscured (including Compton thick) AGN to $\sim$20\% (Hopkins et
al. 2009).

A number of wide area, multi-wavelength surveys with sufficient depth
to view the full AGN population are currently in progress, including:
ECDFS (Lehmer et al. 2005, 0.3 deg$^2$),
OPTX (Trouille et al. 2008, 1 deg$^2$ non-contiguous), 
SWIRE/\chandra\ (this paper, 0.7 deg$^2$), 
AEGIS-X (Laird et al. 2009, 0.67 deg$^2$), and 
C-COSMOS (Elvis et al. 2009, 0.5+0.4 deg$^2$). They are beginning
to provide significant samples of the relatively rare sources,
such as the high redshift or obscured AGN, found in small
numbers in the deep \chandra\ surveys (Luo et al. 2008, Alexander et
al. 2003).
These various surveys will allow us to 
properly characterize these populations and understand their significance to
AGN as a whole.

In this paper we present the 0.7 \sqdeg~SWIRE/\chandra\ X-ray
source catalog, SWIRE identifications and IR, optical and radio fluxes. 
IR properties of X-ray
sources range from the power-law shape characteristic of AGN-dominated
sources, mostly unobscured Type
1 AGN, to SEDs which are dominated by star formation or
host-galaxy emission, implying obscuration of the AGN at these wavelengths
(Franceschini et al. 2005,
Barmby et al. 2006, Polletta et al. 2007, Feruglio et al. 2008, Cardamone et
al. 2008, Gorjian et al. 2008). X-ray hardness is loosely related
to the optical/IR colors, 
generally supporting this view. We characterize the
X-ray and multi-wavelength properties of the SWIRE/\chandra\
sample and take an initial
look at these properties as a function of X-ray hardness and radio-loudness.
Detailed study and modeling of the spectral energy distributions
(SEDs) will be 
presented in a companion paper (Polletta \etal\, in prep.).

\section{X-ray Data and Analysis}
\label{sec:X}
\begin{figure}
\epsscale{1.0}
\plottwo{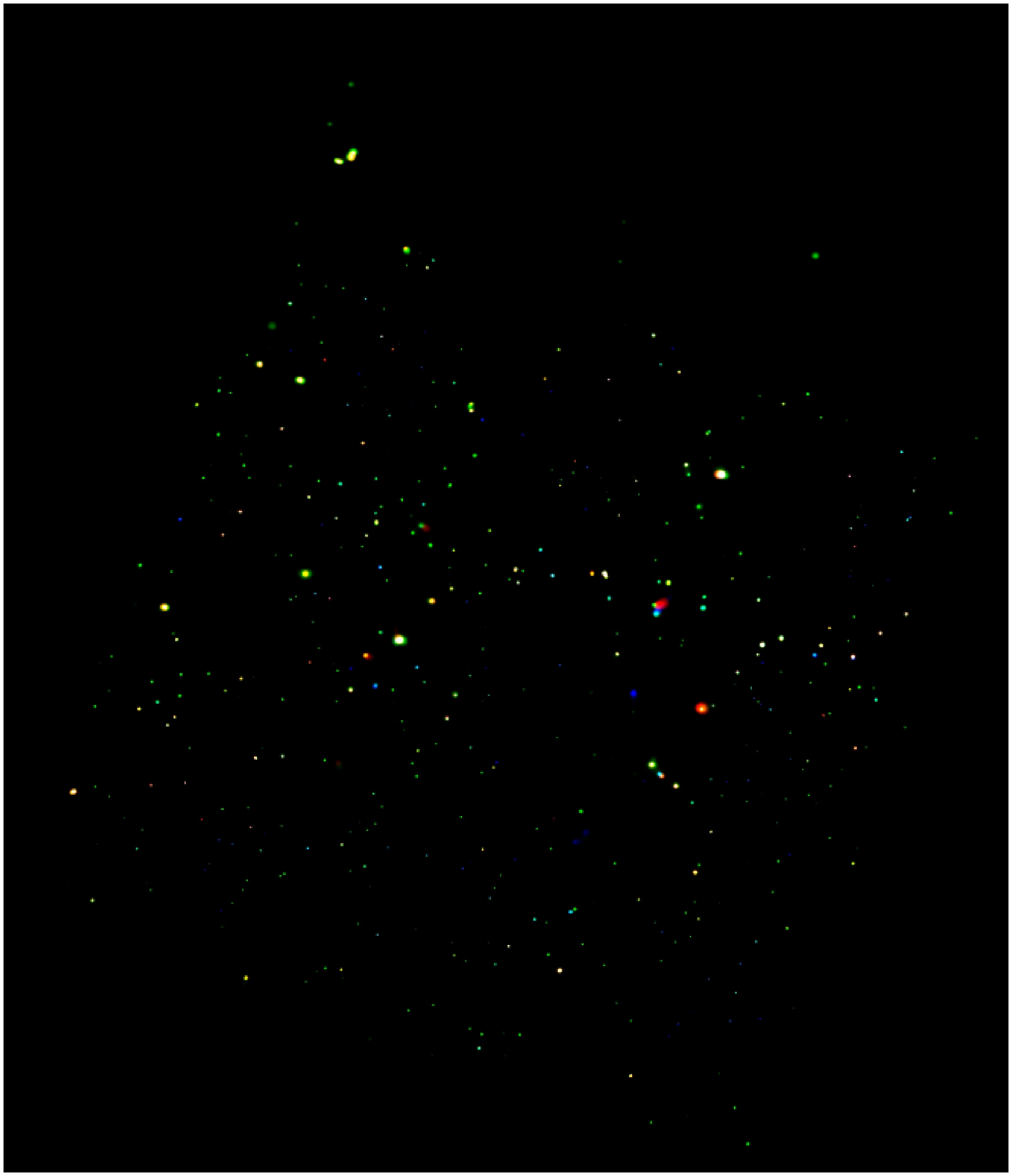}{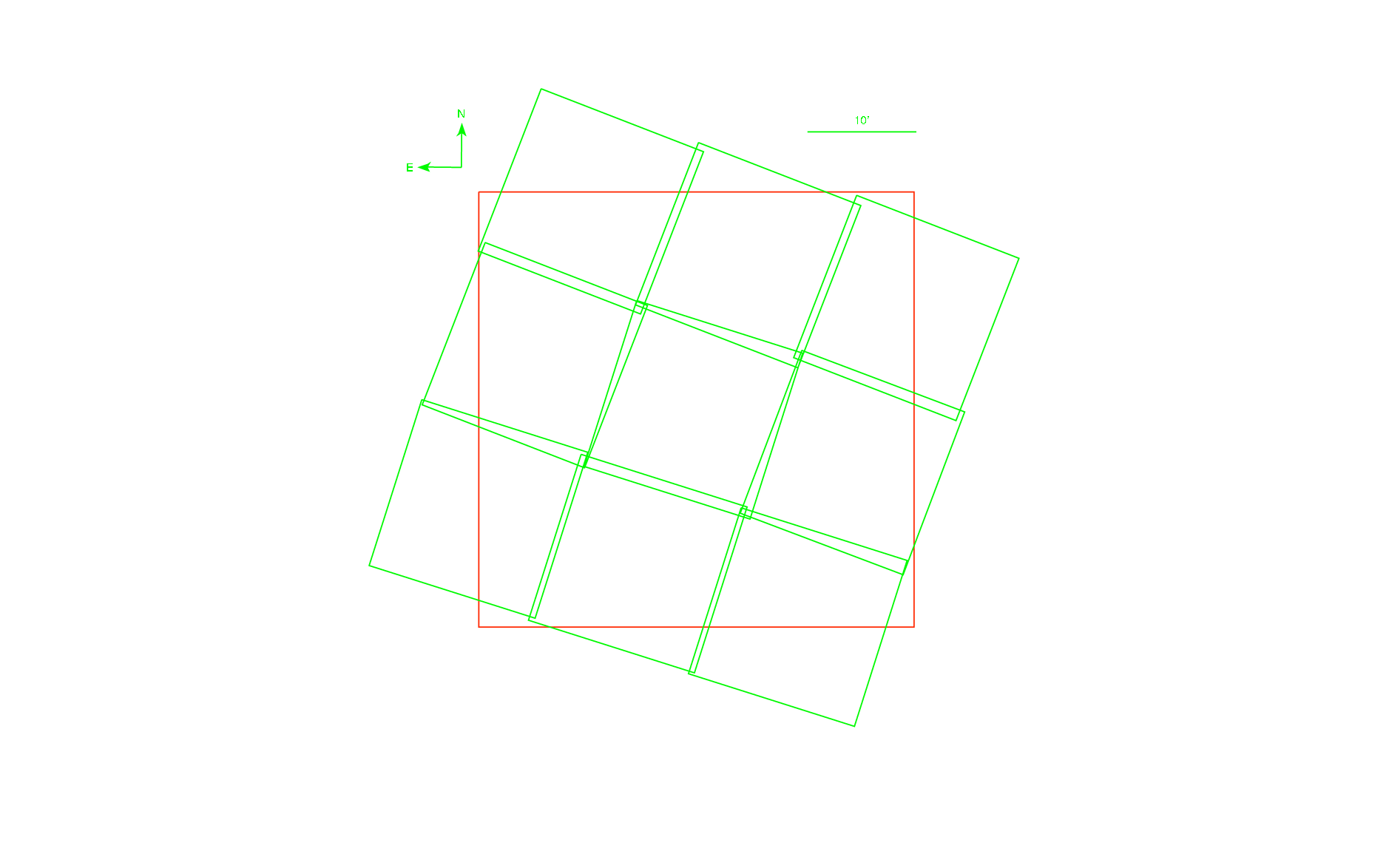}
\caption{Left: Image of the full, mosaic'd X-ray field with counts  
color-coded by energy as follows: red (0.3-0.9keV), 
green (0.9-2.5keV) and blue (2.5-8keV) so that 
soft sources are red and hard ones are blue. Note that sources
appearing extended are generally at higher off-axis angles where 
the spatial resolution is lower; 
Right: Fields-of-view of the \chandra\
observations and the deepest VLA data.}
\label{fg:image}
\end{figure}

We have obtained \chandra\ ACIS-I observations in a 3$\times$ 3
raster of a contiguous 0.7 \sqdeg~area of sky
within the Lockman Hole region of the SWIRE survey, see image in
Figure~\ref{fg:image} (left)
color-coded for X-ray energy.
In addition to the standard 4 ACIS-I chips, we also include the ACIS-S2
chip in our analysis.  The exposure time
for each observation was $\sim$70 ksecs, reaching broad band
fluxes of $\sim 4 \times 10^{-16}$ \fcgs\ (for $\sim 5$ counts,
on-axis). The nine observations, listed in Table~\ref{tab:swireobs}
along with the Galactic $\rm N_H$ values from Stark et al. (1992),
were obtained between 12 and 26 September 2004. They were arranged in a
3$\times$3 raster spaced by 15$'$ in order to ensure overlap of the
individual pointings. The observations
were made within a 15-day period with roll angles (Table~\ref{tab:swireobs})
from 17.60$-$21.20\deg, a sufficiently small range that the fields
overlapped and coverage was contiguous (Figure~\ref{fg:image},
right). 

The data were processed using the XPIPE
pipeline developed for analysis of \chandra\ data for the ChaMP
using CIAO v3.1 (Kim et al. 2004, hereafter DK04). 
XPIPE screens bad data, corrects instrumental effects remaining after
the standard pipeline processing, detects the X-ray sources (using
{\it wavdetect} in the broad band 0.3$-$8 keV) and determines source properties
including counts, fluxes, hardness ratios and colors using the
soft (0.3$-$2.5keV), hard (2.5$-$8.0keV) and broad (0.3$-$8.0keV)
bands as well as the commonly used soft (0.5$-$2.0keV) and hard (2$-$8keV)
(Table 2). 
A threshold of 10$^{-6}$ in {\it wavdetect} was used to accept a
source, corresponding to $< 1$ spurious source per ACIS-I chip, or 4
per field. To avoid finding spurious sources located at the edge of the CCD chips,
a minimum of 10\% of the on-axis exposure was required for source detection.
Exposure maps of the SWIRE/Chandra fields were generated for each CCD
at an energy of 1.5 keV with an appropriate aspect
histogram\footnote{See http://cxc.harvard.edu/ciao/threads/expmap\_acis\_single.}.
The maximum value of the exposure maps is $\sim 600$ cm$^2$ counts/photon
in each CCD. Since the area with exposure map values $< 10$\% of the maximum in
each CCD is $\sim 1$\% of the geometrical area of each CCD, $\sim
~1$\% of the total number of detected sources met this criterion, 
assuming the sources are randomly distributed across the CCD.

A total of 897 sources were detected in this initial
analysis.  From this list, 80 sources were multiply-detected ``duplicate"
sources found in the regions of overlap between observations.  In these cases,
we excluded detections with lower source significance from
the catalog; given that the sources are likely not variable on
the timescale between observations, the lower significance is either due to
larger off-axis angle, lower net counts due to telescope dither when
close to a chip edge or
a combination of the two effects.  An additional 42 sources were deemed
to be spurious based on visual inspection of the X-ray, optical and IR images,
consistent with the
{\it wavdetect} threshold of $\sim$ 4 spurious sources per field.  The final list
(Table~\ref{tab:swirex}) 
thus contains 775 unique X-ray sources, 763 of which are found in the contiguous field
with the remaining 12 detected on the S2 chips in the 3 most northerly pointings.
The X-ray sources, including positions, 
counts, fluxes and hardness ratios 
are listed in Tables~\ref{tab:swirex},\ref{tab:flux}.\footnote{We present
a page of each table here and the
full tables in the online version of the journal and on our website
(http://swire.ipac.caltech.edu/swire/swire.html). SWIRE/\chandra\ 
database access with a search capability is
available at http://cosmosdb.iasf-milano.inaf.it/CHANDRA-SWIRE/index.jsp.
} 
Hardness ratios (defined as (H-S)/(H+S), where H and S are the
counts in the hard (2.5-8.0 keV) and soft (0.3-2.5 keV) bands
respectively) were determined with a Bayesian approach using the 
BEHR program\footnote{See
  http://hea-www.harvard.edu/AstroStat/BEHR (Park et
al. 2006)}. This method is designed for low count
sources where a Poisson distribution is appropriate and
uses as input the hard and soft source and
background counts. It was applied as described in Kim et al. (2007a)
to estimate the hardness ratio and its error 
for all sources detected in the broad band.
To estimate the flux of the X-ray point sources, we first calculated
the energy conversion factor (ECF), which converts 
count rate to flux, assuming a single power law model
for the X-ray source spectrum with photon indices of $\Gamma$=1.2, 1.4,
and 1.7 (see Section 3.2.4 of Kim et al. 2007a (hereafter MK07) for
a detailed description of estimating ECF).
The fluxes presented in Table~\ref{tab:flux} 
assume $\Gamma = 1.7$. For soft/hard-band fluxes where the net counts
were negative, no flux estimation is provided. 
The logN-logS relation (Section 4) was derived adopting the ECFs for 
photon indices of $\Gamma$=1.4 and 1.7.

\begin{figure}
\epsscale{1.0}
\plotone{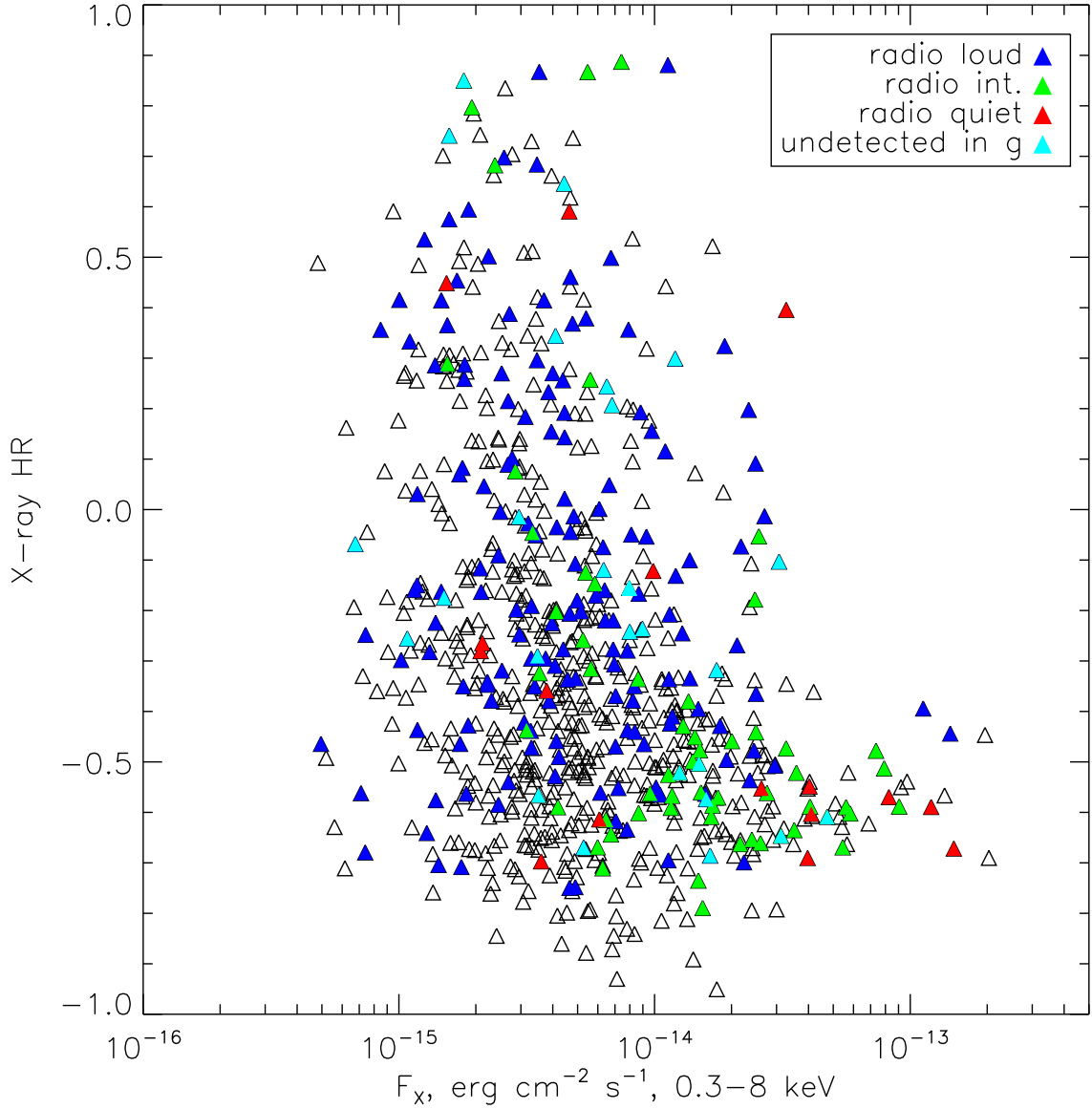}
\vspace{0.5in}
\caption{X-ray hardness ratio as a function of the broad band
X-ray flux. The hardness ratio is determined using a Bayesian approach
which includes estimates for all targets detected in the broad band.
The well-known tendency for hard sources to have fainter X-ray flux is
visible.
Color coded labels show radio-loud, -intermediate and -quiet sources,
classified based on R$_L$ (Section~\ref{sec:rl}). The light blue
are radio-loud sources with  no g$'$ detection. The remainder (open
triangles) are
unobserved/undetected in the radio.
}
\label{fg:XvsHR}
\end{figure}

\begin{figure}
\epsscale{1.0}
\plottwo{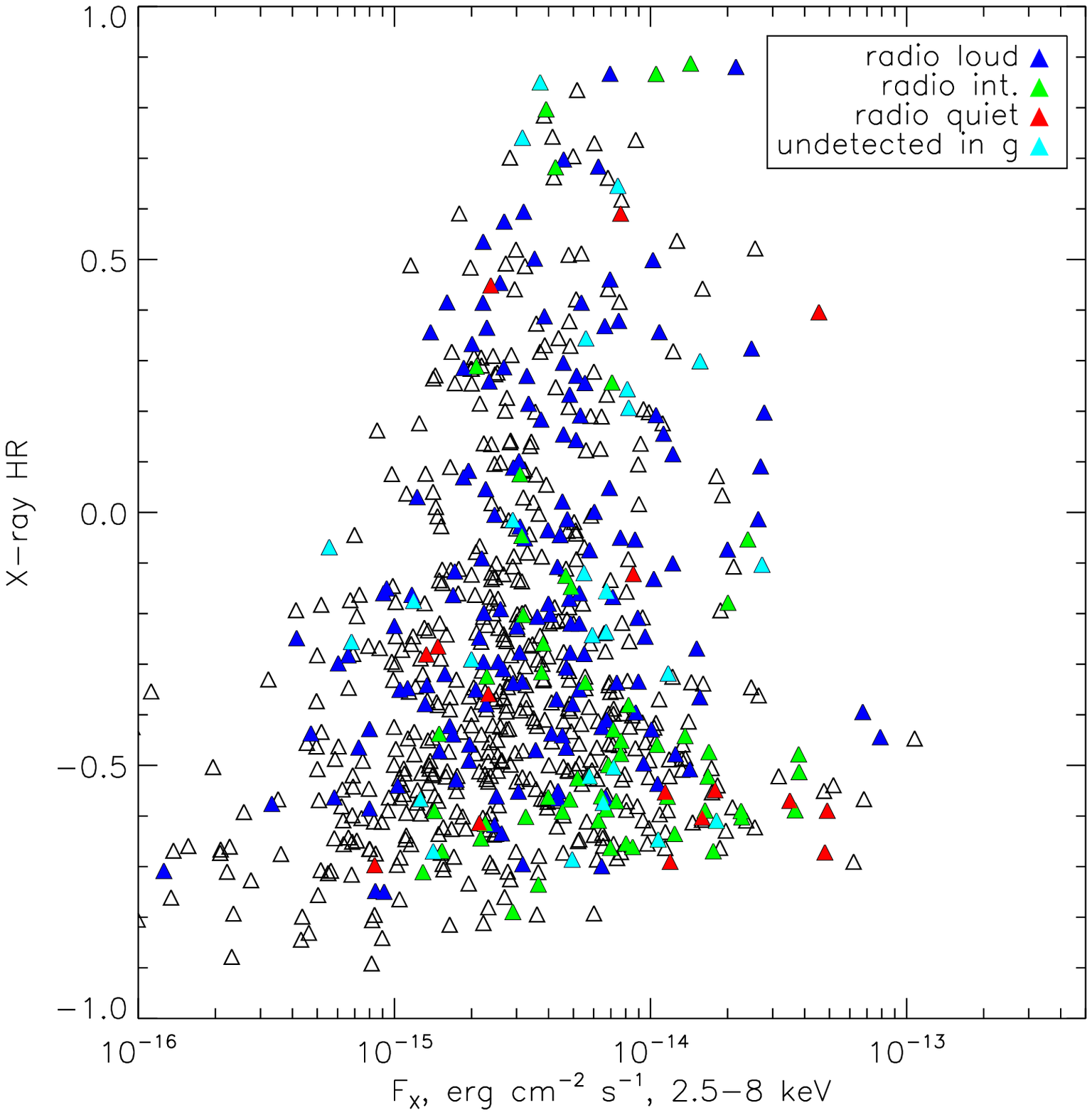}{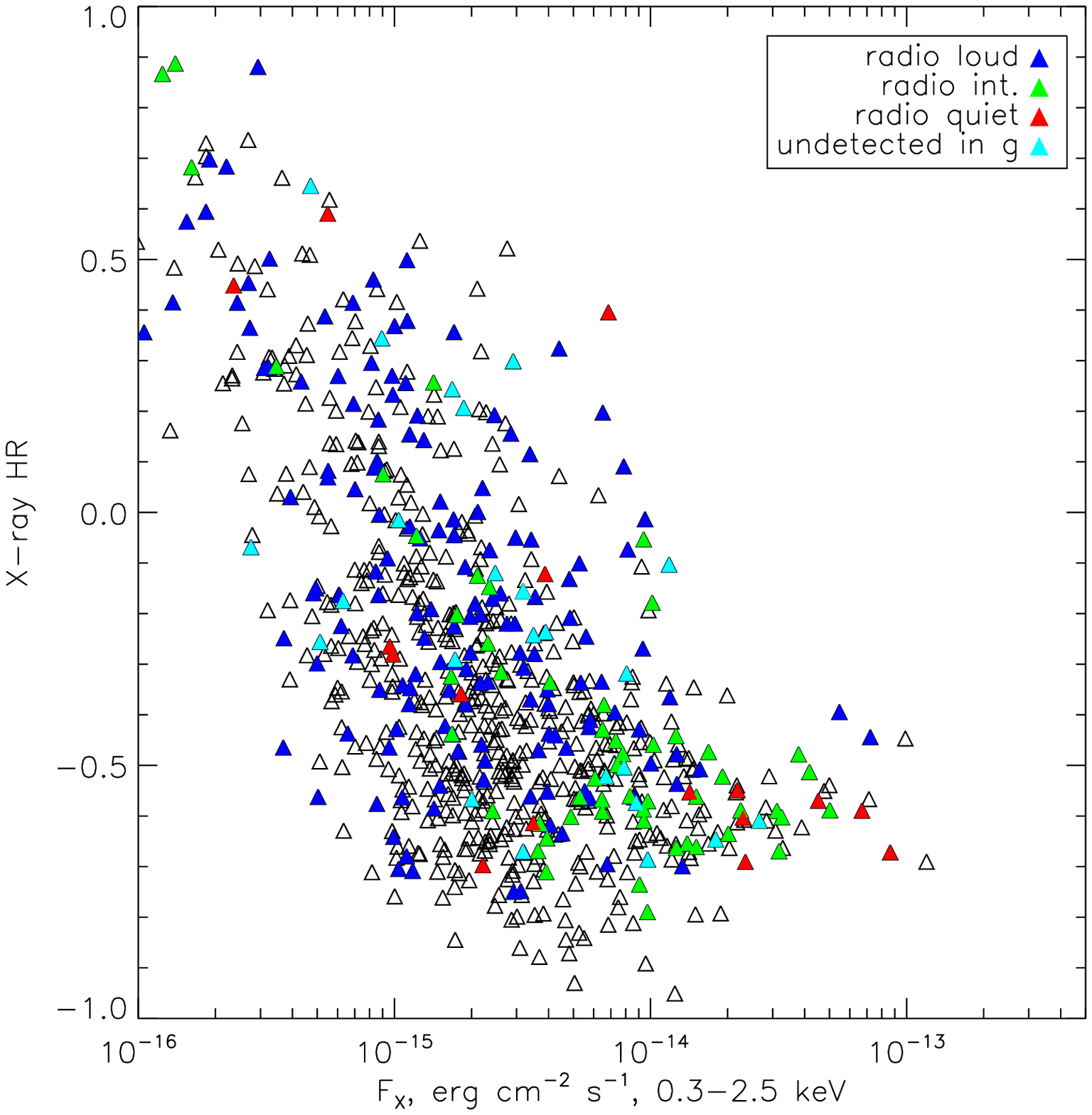}
\caption{X-ray hardness ratio as a function of the hard (left) and soft
(right) band X-ray flux for all the X-ray 
sources which are detected in the broad band.
Color coded labels are as in Figure~\ref{fg:XvsHR}.
The tendency for hard sources to have fainter X-ray flux is
strongly visible in the soft band. Its absence in the hard band
confirms the usual interpretation that soft X-ray absorption 
reduces the soft and broad band fluxes, causing harder spectra and lower
observed fluxes in these sources.
}
\label{fg:XvsHR_2}
\end{figure}

The X-ray flux limit of the sample is F(0.3-8keV) $\sim 4 \times 10^{-16}$
\fcgs (5 counts, on-axis). 
The brightest source is at 3$\times 10^{-13}$
\fcgs . This survey covers the flux range in between the deepest
\chandra\ surveys and the wide-area, shallow surveys. Figure~\ref{fg:XvsHR}
shows the X-ray hardness ratio 
as a function of flux in the broad band. The usual trend for hard
sources to have fainter fluxes is visible. Figure~\ref{fg:XvsHR_2}
repeats Figure~\ref{fg:XvsHR} as a function of the hard (left) and soft (right)
band fluxes. The trend is stronger in the soft band
and largely absent in the hard band, 
consistent with the general interpretation of the hardness being due to
obscuration (DK04, MK07).
Radio-loud, -intermediate, and -quiet sources are also labelled, based
on the original R$_L$ classification scheme (see Section~\ref{sec:rl}
for more discussion). Radio-loudness does not show any particular trend
in these figures.

During the final stages of preparation of this paper, 
an X-ray source list derived from these \chandra\ data was published
(CLANS, Trouille et al. 2008). 
This source list includes 761 X-ray sources, similar to
our 775, but a detailed comparison yields significant differences.
The CLANS list was derived using a 
{\it wavdetect} threshold of $10^{-7}$. 
We find 94 CLANS sources with no match in our X-ray source list. Ten of these
were in our preliminary list but were deleted as spurious following visual
inspection. There are 113 sources in our X-ray list which have no CLANS
match, including
33 sources with $\alpha > $162.2\deg , a sky region not included
in their analysis. With the exception of these 
33, the sources which appear in only one list are at
low flux levels. 
There are 662 sources in common, 5 of which are counted as double in
the CLANS list but confirmed by visual inspection to be single in
ours. The derived X-ray fluxes in the two samples generally agree well, 
except at low
count rates in the hard band. 


\subsection{X-ray Spectral Analysis}

Fifteen of the X-ray sources have $>$300 net counts, facilitating spectral analysis.
Spectra and associated response functions were extracted and the spectra were
grouped with a minimum of 15 counts per bin.  Fits were performed utilizing
the {\it CIAO (v3.4)} tool {\it Sherpa} with the Levenberg-Marquardt
optimization method and $\chi^2$ statistics with variance computed from
the (chi-dvar).  90\% errors were computed using the {\it projection} method
in {\it Sherpa}.

\begin{figure}
\epsscale{0.5}
\plotone{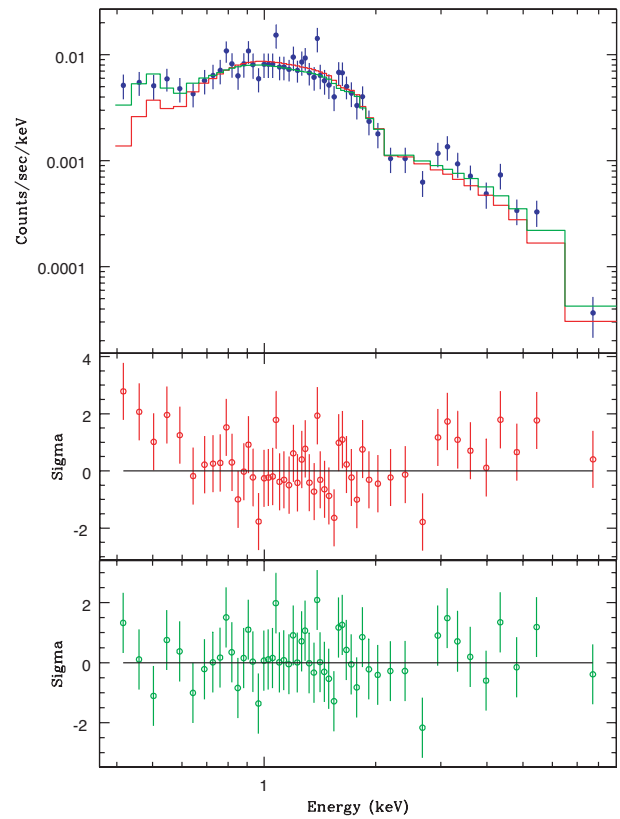}
\caption{X-ray spectrum of the source CXOSW J104540.1+584254. 
Plotted in red is the best single
power-law fit and in green a fit utilizing two power-law components, the simplest
possible model to account for the clear soft excess in the spectrum.
The residuals to both fits are shown underneath using the same color-coding.
}
\label{fg:soft-excess}
\end{figure}

We first fit all the spectra with a power law model
(F$_{\nu}= \nu ^{-\alpha_e}$, where \alpe~is the energy index and
photon index, $\Gamma = 1+\alpha_e$)
and cold absorption at redshift zero (\nh ). The minimum allowed \nh\ was set
to 5.9$\rm \times\ 10^{19}\ cm^{-2}$, $\sim 1 \sigma$ below the lowest
Galactic value (Table~\ref{tab:swireobs}, Stark et al. 1992).
The results are shown in Table~\ref{tab:xspec}.
Eleven sources are acceptably fit
with this model alone. The average power-law slope,
$\Gamma$ = 1.89$\pm$0.36, is consistent with the results for luminous sources in
the \chandra\ Deep Field South (CDFS, Tozzi et al. 2006).  For 5 of
these 11
sources, \nh\ hit the minimum allowed value, suggesting a soft excess in
the spectrum that was not accounted for in the model.
To examine the soft excess in more detail, we performed fits
with an absorption component frozen to the Galactic value and added a
soft excess component, modelled with a power-law,
to the spectrum.  Since the single power-law models
provided statistically acceptable fits, we cannot substantially improve upon
them with the additional soft component. In 4 cases the
additional soft component provides a visually more acceptable fit in the
low energy range.  For the fifth, 
CXOSW J104540.1+584254, 
while the single power-law fit is statistically
acceptable ($\chi^2$ = 59.50 for 50 degrees of freedom), the addition
of a soft component does improve the fit.  The simplest possible model is the
addition of a second power-law component, which results in slopes
of $\Gamma_1$ = 5.00, $\Gamma_2$ = 1.89, with $\chi^2$ = 46.63 for
48 degrees of freedom  (see Figure~\ref{fg:soft-excess}).

Of the remaining 4 sources, only 
one: CXOSW J104803.4+585547, 
has a statistically acceptable fit to
a Mewe-Kaastra-Liedahl thermal
plasma model (MEKAL) with temperature kT = 3.15 kV.  It is likely that
poor counting statistics prevent us from obtaining
good model fits for the remaining 3 sources.

\subsection{Emission Line Feature in CXOSW J104404.0+590241} 

\begin{figure}
\epsscale{0.5}
\plotone{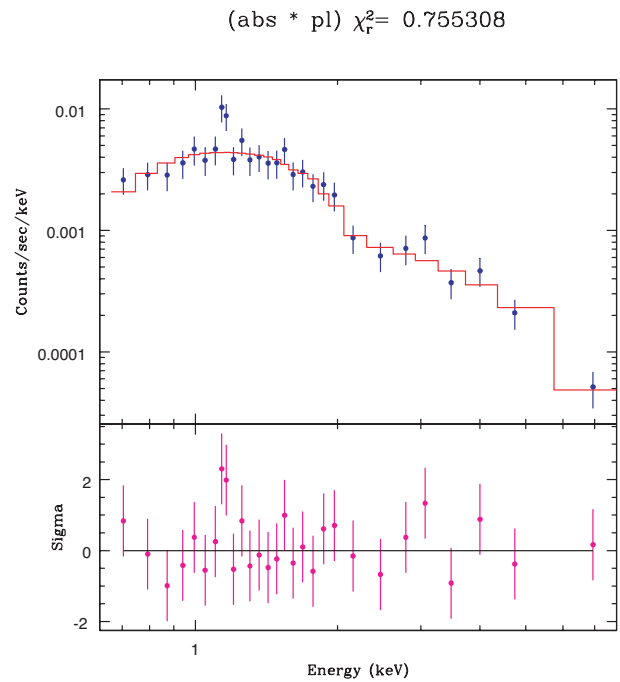}
\caption{X-ray spectrum of the source CXOSW J104404.0+590241 
with best-fit single
power-law model.  The residuals are shown below.
There is the suggestion of a spectral line at $\sim$1.14 keV
identified with Si K$\alpha$ at a redshift $\sim 0.5$.
The signal-to-noise is too low at higher energies to confirm (or not) the
presence of an Fe K$\alpha$ line at the same redshift, with energy $\sim 4.2$ keV.
}
\label{fg:xray-line}
\end{figure}

Source CXOSW J104404.0+590241 shows
evidence for a spectral line at 1.14 keV (see
Figure~\ref{fg:xray-line}).  It is not required for the fit, but is preferred
at around the 1.5 $\sigma$ level based on an F-test,
having an (unconstrained) equivalent width
$\sim 80$ eV. In general a single line is interpreted as
the Fe K$\alpha$ line at 6.4 keV (rest-frame), the strongest emission line
present in AGN X-ray spectra. However, in this case the source would be
at z = 4.6 and the X-ray luminosity would be very high at
4.9$\rm \times\ 10^{46} erg\ s^{-1}$.

Trouille et al. (2008)
give z=0.5 for this source (their \#136) and assign
a BLAGN type based on a strong Balmer series in the optical spectrum 
(Trouille,
private communication).
The closest possible identification for the X-ray line at this redshift is 
the Silicon K$\alpha$ fluoresence line, rest energy 1.74keV,
yielding z = 0.52. This line has been observed in AGN, though not
generally in isolation. For example, photoionized material in the Broad-line
Radio Galaxy, 3C445, provides several emission lines, including Si K$\alpha$
with an equivalent width 73 eV
(Sambruna et al. 2007). At the redshift of this source, the other 
expected lines are redshifted to
energies below the observed band in this source, apart from 
Fe K$\alpha$, $\sim$ 4.2 keV. However 
the signal-to-noise in the higher-energy X-ray spectrum is too low to
confirm or exclude the presence of a Fe K$\alpha$ line 
(Figure~\ref{fg:xray-line}).
We conclude that the emission line is Si K$\alpha$ at z$\sim 0.5$.

\section{Multi-wavelength X-ray Source Identification}
\label{sec:XID}

The extensive multi-wavelength data for the SWIRE/\chandra\ field
are described in Polletta et al. (2006) and will be presented in detail,
along with classification of the sources and SED fitting, in
a companion paper (Polletta et al., in prep.).
For completeness they
are briefly summarized here. The {\it Spitzer} data include
imaging in the 4 IRAC bands to 5$\sigma$ depths of 4.2, 7.5, 46 and 47 $\mu$Jy
at 3.6, 4.5, 5.8, 8.0 $\mu$m respectively. MIPS imaging at 24 $\mu$m
reaches depths of 209 $\mu$Jy ($5 \sigma$, Shupe et al. 2008).
The details of the IRAC and MIPS data
processing are given in Surace (2005).
Here we will refer to the optical and infrared imaging data
in order to characterize
the sample in comparison with other X-ray surveys.
Optical imaging in U, $g'$, $r'$, $i'$, to 5$\sigma$ limiting Vega
magnitudes of 24.8, 25.9, 25.2 and 24.4 respectively, was obtained with
the MOSAIC Camera on the KPNO 4-m Mayall telescope for
the field center and with $\sim 0.5-1$ magnitudes brighter limits towards
the field edge (Polletta et al. 2006). 


The X-ray source list was cross-correlated with the optical
and {\it Spitzer} source lists in search of matches.
The SWIRE catalog contains all 
sources detected in at least one IR band in the Lockman Hole field.
The cross-match procedure searches for sources within a circular
radius corresponding
to the geometric sum of 2\,\arcsec and the X-ray positional error
or 2\,\arcsec, whichever was greater, to ensure no
potential matches were missed.
We find an IR counterpart for
749 X-ray sources 
and 333 have 24 $\mu$m identifications. 
Visual inspection of the images was carried out for 68 sources with multiple associations.
The closest source is preferred when it was more than $\sim 2$ 
times closer than any other. For counterparts at similar distances from the
X-ray source, the brightest in the IR is generally preferred. In some cases 
the source SED is taken into account, preferentially selecting 
the counterpart with 
an AGN-like SED. This inspection led to the selection of 
preferred counterparts in 63 cases while in 
the remaining 5 cases the identification remains ambiguous. 
Further inspection of the 26 sources with no IR counterpart
revealed the presence of an
IR source below the formal detection threshold in 8 cases.
In 10 cases a visible counterpart is not listed in the
catalogs due to the presence of a nearby bright star.
Of the remaining 8 X-ray sources, 4 only have optical identifications and 4 
are genuine X-ray only sources for which the optical and IR
emission is fainter than the SWIRE sensitivity limits. Thus we have 
identified an optical/IR counterpart for 771 X-ray sources, or $>$99\%
of the sample.
Given the uncertain IDs, contaminating sources and very faint flux
levels of some of the identifications, we have
accurate optical and/or IR flux measurements for 744 X-ray sources,
corresponding to 96\% of the entire X-ray sample (Table~\ref{tab:swireID}).
Those sources with no optical/IR counterpart, those with ambiguous IDs
and those with contaminating sources which prevent count extraction are
labelled in Table~\ref{tab:swireID} (electronic version).


A deep 1.4 GHz radio map covering the central 40$' \times 40'$ of the
SWIRE/\chandra\ field was obtained at the VLA (Very Large Array). 
Owen \& Morrison
(2008) present a 5$\sigma$ catalog for which the flux limit
in the center of the radio field is 15 $\mu$Jy and a factor of $\sim
5$ brighter by 20$'$ off-axis (Owen \& Morrison 2008).
Shallower data are available for the full \chandra\ field-of-view.
In this paper we
use a radio source list extending down to a deeper, 4$\sigma$, flux
limit, reliable 
given the additional
requirement of identification with an X-ray source. This yields a 
radio flux limit (4$\sigma$) of 11 $\mu$Jy at the field center.
The radio source list was cross-correlated with the
X-ray source list to determine candidate radio
identifications. Candidate matches were determined based upon the positional
errors in both radio (which take into account extent) 
and X-ray and then confirmed by visual inspection. The low
density of radio and X-ray sources results in no ambiguous matches.
Radio sources were identified with 251 X-ray sources in the overlapping field
of view (Figure~\ref{fg:image}, right). 
We take a first look at the classification of the radio-detected sources
in this paper. A detailed analysis of the radio properties will be presented in a
future paper.

\subsection{Extended X-ray Sources}

We searched for extended X-ray sources using {\it vtpdetect}.
Two candidates were found: CXOSW J104740.7+590704 
and CXOSW J104653.1+592648. 
Modelling of their point spread function (PSF) at 1 keV confirmed their extent
at 95\% significance.

CXOSW J104740.7+590704 appears to be associated with a group of galaxies in the optical and IR images.
The central galaxy is at a redshift of z=0.354.  The X-ray emission is soft (HR = -0.22) and can
be fit with a simple power-law of slope $\Gamma$ = 2.25.  The extent of the X-ray emission
is almost an arcminute in radius, though the association of galaxies is smaller.
The 0.3-8 keV flux is $\rm 1.0\times 10^{-13}\ erg\ cm^{-2}\ s^{-1}$ which, at the redshift
of the central galaxy, corresponds to a luminosity of $\rm 4.5\times 10^{43}\
erg\ s^{-1}$.
Though the X-ray emission does not appear to contain any discrete
sources, it remains possible
that one or more AGN contribute.

CXOSW J104653.1+592648 is associated with two galaxies in the IR which cannot be visually separated
in the optical.  The diffuse X-ray emission is concentrated in a much smaller area
than the previous source: 
(30\arcsec\ in radius) which, given the source's large off-axis angle (7.3\arcmin),
suggests that the diffuse emission is likely associated with these two galaxies only.
The X-ray emission is quite soft (HR = -0.43) and the spectrum is well-fit by a soft
power-law with absorption and photon index, $\Gamma$ = 3.0.  The unabsorbed X-ray
flux is $\rm 1.3\times 10^{-13}\ erg\ cm^{-2}\ s^{-1}$. A spectroscopic
redshift has not been measured for either IR-identified galaxy
so the luminosity cannot be determined.
No discrete sources are detected in the X-ray emission which is centered on
the gap between the galaxies, so, though still possible, it is
unlikely to be dominated by an AGN

\section{The Log N vs. Log S Relation}
To determine an accurate logN-logS relation, it is necessary to correct
for the incompleteness of the sample as well as for instrumental
effects such as vignetting and the off-axis degradation of the PSF.
The actual source detection probability in a $Chandra$ field is a complex function
of off-axis angle and source counts: the detection probability
decreases as off-axis angle increases and as source counts decrease (e.g. MK07).
Therefore, to accurately determine the sky coverage of the SWIRE/\chandra\ sample,
we performed a series of Monte Carlo simulations to correct
incompleteness and biases within the sample fields.
The technique for the Monte-Carlo simulations is based on previous studies
(e.g. Kim and Fabbiano 2003; MK07; Kim et al. 2007b) and consists of three steps:
(1) generating artificial X-ray sources with
MARX\footnote{See http://space.mit.edu/CXC/MARX/ and MARX 4.0 Technical
Manual.},
(2) adding them to the original observed image without subtracting any real,
observed X-ray sources, and (3) detecting
these artificial sources with $wavdetect$ and extracting source
properties. 
In step (2), we used the real \chandra\ observations
to accurately reflect the effects of background counts and source 
confusion/crowding in the SWIRE fields.

We performed simulations using all 4 CCDs in all 9 
observations (see \S 2), generating
$1,000$ artificial X-ray sources in each of our 9 sample
fields extending to a flux level 10\% of the survey detection limit. 
This corresponds to $\sim13,000$ artificial X-ray sources per $\rm
deg^{2}$. The number of artificial sources detected in each field depends on
the effective exposure time of the observation and
the neutral hydrogen column density, $N_{H}$,
toward the observed region of the sky.

The observed X-ray differential logN-logS  is described by a
broken/double power law
with a faint slope of $\sim-1.5$ and a bright slope of $\sim-2.5$
(Yang et al. 2004, Basilakos et al. 2005, Chiappetti et al. 2005)
in most energy bands; however, the break flux has not been well determined.
In our simulations, we assumed a cumulative logN-logS distribution with a
single power law of slope of $-1$, corresponding to
a slope of $-2$ in the differential logN-logS. This is
the average of the faint and bright slopes
from the literature, in the 0.3-8 keV band.
Since we use the fraction of artifical sources
detected as a function of flux to determine the 
sensitivity (Vikhlinin et al. 1995, Kim \& Fabbiano 2003),
the exact form of the assumed logN-logS distribution
is not critical. From the assumed logN-logS distribution,
we selected artificial source fluxes and placed them in the
actual event files at random positions, taking care not to over-crowd each CCD
chip (see below). Applying the same detection software to these events
files as we did to the original observations, we found that on 
average $24\%$ of the $9,000$ artificial
X-ray sources are detected in our simulations. 
The flux range of the detected, artificial sources: $5\times10^{-16}$
to $\rm 5\times10^{-10}$ \fcgs\ in the broad band
(Table~\ref{tab:bands}); covers the flux
range of the observed SWIRE/\chandra\ X-ray point sources ($ 4\times10^{-16}$
to $2\times10^{-13}$ \fcgs ).
The result was a total of $2,127$ artificial X-ray sources in the 9
SWIRE/\chandra\ fields. At 2.8 times the 775 observed sources,
this is statistically sufficient to derive the sky coverage 
as a function of flux and to correct the effects of 
incompleteness and bias on
the logN-logS relations derived for the SWIRE/\chandra\ sample.

The spectrum of the artificial sources was assumed to
be a power law. 
Observed photon indices in this flux range generally span 
$\Gamma=1.5-2$ (KD04, MK07).
In addition,
Tozzi et al. (2006) performed X-ray spectral analysis for 82 X-ray bright
sources in the CDFS, and found a weighted mean value for
the slope of the power law spectrum, $<\Gamma>\simeq1.75\pm0.02$,
and no significant correlation between the photon index,
$\Gamma$, and the intrinsic absorption column density $N_{H,int}$.
The flux range of these bright sources in the CDFS overlaps 
the faint flux end of the SWIRE/\chandra\ sources. We therefore
assumed that the SWIRE/\chandra\ sources have a photon index,
$\Gamma\sim1.7$.
We assumed a Galactic absorption, $N_{H}$ (Stark et al. 1992), 
for each observation; we did not include
intrinsic absorption for the artificial source spectra.
The spectrum of each X-ray point source was generated
using the XSPEC\footnote{See http://xspec.gsfc.nasa.gov/.} package.

The artificial source's position was randomly selected in each CCD chip area,
but was rejected if the source area at a selected position had
an exposure map value of less than $10\%$
of the maximum.
This requirement is identical to that in the SWIRE/\chandra\ X-ray point
source catalog reduction procedure.
To avoid over-crowding of the artificial sources,
the $\sim250$ artificial sources per CCD were
divided into several groups to be added into the observed image:
while we did not allow the artificial X-ray point sources to overlap
one another,
we allowed overlap between artificial and real X-ray
sources to provide an estimate of source confusion in each observed field.
This resulted in $\sim10$ simulated images per ACIS-I
CCD, corresponding $\sim360$ CCD images (event files) on which we ran
$wavdetect$ ($xapphot$).
Since $\sim24\%$ of the artificial sources ($\sim2100$) are detected,
on average we added $\sim6$ detectable, artificial sources
to each simulated image.
The net counts of those artificial sources which overlap/blend 
with real sources were
corrected following the methods described in
\S3.2.2 of MK07.

Using the simulation results, we derived the sky coverage 
as a function of flux in 6 energy bands (Table~\ref{tab:bands}).
The sky coverage area is the fraction of artificial sources detected 
at a given flux, multiplied by the total sky area,
and it is used to correct the incompleteness and bias in  
the derived logN-logS relations.
The full sky area of the SWIRE/\chandra\ is $0.7~deg^{2}$.
The geometrical area of a $Chandra$ CCD chip is $0.0196~deg^{2}$; however,
the net effective area is slightly larger due to the dither.
To accurately calculate the effective area,
we follow the same method as $xapphot$:
all pixels in the exposure map were summed,
excluding those pixels with
an exposure map value less than $10\%$ of the maximum
within the corresponding source area.
This criterion automatically excludes pixel positions
located near the edge of the CCD chip.
We note that our sample is complete at the $90\%$ level at a flux of
$1.6\times10^{14}$ \fcgs\
in the $0.5-8$ keV when $\Gamma=1.4$ is assumed.

The cumulative logN-logS relation for sources brighter than a given flux $S$,
corrected by the corresponding sky coverage at $S$, is:
\begin{equation}
N(>S)=\sum_{S_{i}>S} {1 \over \Omega_{i}}, 
\end{equation}
where $S_{i}$ is the flux of the $i$th X-ray point source and $\Omega_{i}$
is the sky coverage, that is the maximum solid angle over which a
source with flux $S_{i}$ is detectable.
Using the SWIRE/\chandra\ X-ray point sources
and the corresponding sky coverage,
we derived the cumulative logN-logS relations for the SWIRE/\chandra\
X-ray point sources.
Since the differential logN-logS relation is a derivative form of the
cumulative logN-logS relation,
we derived it from the cumulative logN-logS relation
resulting from equation (1) as follows:
\begin{equation}
{\left.{dN \over dS}\right|}_i 
=-{{N_{i+1}-N_{i}} \over {S_{i+1}-S_{i}}}
\end{equation}
where $N_{i}$ is the cumulative source number at flux $S_{i}$.
Since the sky coverage rapidly decreases near the faint flux limit,
there are large statistical errors for the logN-logS distribution
at faint fluxes.
Thus, for better statistics, we present the logN-logS relations brighter
than the flux corresponding to
$10\%$ of the full sky coverage.
In Figure 6,
we display the SWIRE/\chandra\ differential ($left~panels$)
and cumulative ($right panels$) logN-logS relations in 3 energy bands.
Statistical errors on the logN-logS
are assigned following Gehrels et al. (1986).

We fitted the differential logN-logS with a broken power law
as follows:
\begin{equation}
{{dN} \over {dS}} = \left\{\begin{array}{ll}
K(S/S_{ref})^{-\gamma_{1}}, & S<S_{b}, \\
 & \\
K(S_{b}/S_{ref})^{(\gamma_{2}-\gamma_{1})}(S/S_{ref})^{-\gamma_{2}}, & S \ge
S_{b}, \\
\end{array}
\right.
\end{equation}
where $K$ is a normalization constant and $S_{ref}$ is a normalization flux.
In this study, we set a normalization flux of $S_{ref}=10^{-15}$
\fcgs , 
$S_{b}$ is the break flux at which the slope of the differential
logN-logS changes, $\gamma_{1}$
and $\gamma_{2}$ are faint and bright power law indices. 
The best fit parameters for the differential logN-logS
for various X-ray energy bands and photon indices of 1.4 and 1.7 
are listed in Table~\ref{tab:lnls}
and displayed in Figure 6. The best fit parameters for the differential
logN-logS of
the ChaMP and ChaMP+CDFs (Kim et al. 2007b) are also listed in Table~\ref{tab:lnls}
for ease of comparison.
The choice of photon index, $\Gamma$, has little effect on 
$\gamma_{1}$ and $\gamma_{2}$, but it shifts $S_b$ somewhat.
In the different energy bands the break flux, S$_b$ has different values,
similar to the results of the ChaMP and the ChaMP+CDFs.
We note that the best fit parameters $\gamma_{1}$ and $\gamma_{2}$
and $S_{b}$
agree, within the uncertainties, with previous studies
covering comparable flux ranges, such as
the ChaMP (Kim et al. 2007b). The faint end slope is somewhat steeper
than that for the combined ChaMP+CDFs (Table~\ref{tab:lnls}) or the results of
Trouille et al. (2008). This is most likely due to the inclusion of the
fainter flux range from the CDFs in the latter, combined surveys.

\section{Multi-wavelength Properties of the X-ray Sources}

In Table~\ref{tab:swireID} 
we present multi-wavelength fluxes for the
SWIRE/\chandra\ X-ray sources along with the SWIRE ID number for the
matched source. We include g$'$, r$'$ optical magnitudes,
3.6$\mu$m and 8.0$\mu$m {\it Spitzer} IRAC fluxes, 24$\mu$m {\it Spitzer} MIPS
fluxes and 20cm (1.4 GHz) VLA radio fluxes. 
The full table is available in electronic form only.

\subsection{X-ray and Optical Properties}
\label{sec:rX}
Figure~\ref{fg:rX} shows r$'$ magnitude vs. broad band X-ray
flux for the SWIRE/\chandra\ X-ray sample.
The f$_x$/f$_r$=0.1$-$10 region defined to include most
extragalactic sources (most of which are AGN)
in the Einstein Medium
Sensitivity Survey (EMSS, Stocke \etal\ 1991, hereafter ``the AGN region")
is shown by black dashed lines.
X-ray sources identified with extended optical sources and those with no
IR counterpart are labelled. Sources are colored according to their X-ray
hardness ratios: hard (blue), medium (green) and soft (red). A hardness
ratio, HR$\sim 0$, implies an equivalent hydrogen absorption column
density, \nh~$\sim 10^{22}$ cm$^{-2}$ for a typical AGN power law spectrum.

Figure~\ref{fg:rX} also includes tracks for
optically-bright, Type 1 (Elvis et al. 1994, green lines) and near-IR red
(Kuraszkiewicz et al. 2009a, red lines) AGN SEDs,
based on these low-redshift (z \lax 0.4) samples,
as a function of redshift up to z=4 (at the lower flux end).
The 90\% envelopes,
shown as dotted lines, include the range in shape around this median, but
do not include the range in luminosity expected at a given redshift. 
The tracks for the AGN SEDs are computed over a redshift range from
z=0.025 to z=4 in increments of z = 0.025.  At each redshift interval,
the luminosity distance corresponding to the redshift is computed.
This distance is used to compute the observed flux for an object at
that redshift in each of the bands plotted.  No
absorption is assumed.  For the X-ray flux, we use a monochromatic
flux density at 1 keV assuming a power-law spectrum with slope of
$\Gamma = 1.7$. The 1 keV luminosities (log) of the two medians are 44.61
(Elvis et al. 1994) and 43.51 (Kuraszkiewicz et al. 2009a).
We smoothed the Elvis et al. (1994) $\pm 90$\% envelopes slightly
in the radio range to remove artifacts due to poor data sampling.

As is typical, most of the X-ray sources lie within the range expected for
extragalactic sources, which are predominantly AGN
(f$_x$/f$_r$=0.1$-$10, ``AGN region"). Below this region, where
f$_x$/f$_r< 0.1$, there are 102 sources, 13\% of the sample.
The Elvis et al.
(1994) median as a function of redshift (green lines) predicts
AGN in a small area at the bright end of the AGN region in Figure~\ref{fg:rX},
demonstrating the particular
subset of the luminous AGN population targeted in traditional blue, optical
surveys. The red AGN median (Kuraszkiewicz et al. 2009a, red lines) covers
a much broader part of the AGN region, extending to fainter fluxes
(Figure~\ref{fg:rX}) and suggesting 
that faint sources are predominantly red and/or high redshift.
Red optical colors may be due to a combination of one or more 
factors including: Eddington ratio (L/L$_{Edd}$),
obscuration, host galaxy and scattered light contributions
(Kuraszkiewicz et al. 2009b).
A decrease in L/L$_{Edd}$ decreases the strength of the optical-UV
blue bump (Witt et al. 1997) 
so that sources with low L/L$_{Edd}$ will tend to lie on the optically
fainter side of the AGN region (Kuraszkiewicz et al. 2009b).
Pure obscuration of optically-bright,
Type 1 AGN moves them below the AGN region in Figure~\ref{fg:rX}
into the range of galaxies and XBONGS. Thus 
X-ray hard sources in this region may be low-redshift, obscured AGN. 

The few X-ray sources with no {\it Spitzer} detection
(Section~\ref{sec:XID}) are
distributed throughout Figure~\ref{fg:rX} and so are likely a mixed bag
of source types.

\subsection{X-ray Hardness Ratio}

The X-ray hardness ratio distribution in Figure~\ref{fg:rX} generally
supports the above discussion. The tendency for
hard sources to lie at fainter soft X-ray flux is generally interpreted
as due to obscuration (DK04, MK07, Section~\ref{sec:X}).
At higher redshift, though, the effect
of obscuration at low energies is shifted out of the observed band,
softening the observed spectrum so that obscured, higher redshift sources 
cannot be distinguished by their X-ray hardness. 
Thus sources which are observed as X-ray hard tend
to lie at low redshift (MK07). An investigation of
the relation between optical/IR SEDs and X-ray obscuration requires
redshifts in order to place reasonable limits on the
amount of X-ray absorption for each source and will be presented in the
companion SED paper (Polletta \etal\, in prep.).

Sources at faint flux levels in Figure~\ref{fg:rX}
are preferentially hard. Most of the soft sources lie in 
the AGN region. Those sources which are
optically-bright and X-ray faint lie below the AGN region in Figure~\ref{fg:rX},
are X-ray medium/hard and optically extended,
consistent with being low-redshift galaxies or obscured, Type 1 AGN.

\subsection{X-ray and mid-IR Properties}
\label{sec:5.3}

Figure~\ref{fg:IRvsX_HR} shows the broad band X-ray vs 3.6 $\mu$m, 4.5
$\mu$m, 5.8 $\mu$m and 8.0 $\mu$m fluxes with the tracks for median
optically-selected, Type 1 (green) and red (red) AGN SEDs superposed.
Optically extended
X-ray sources generally lie at higher IR fluxes than the AGN in these
figures, due to the lower observed X-ray flux of
galaxies or low-luminosity/absorbed AGN.
Stronger IR flux, due for example to larger amounts of heated
dust, may also contribute. 

There is good separation of the Type 1 and red AGN SED tracks in
Figure~\ref{fg:IRvsX_HR}. It is
clearer than in Figure~\ref{fg:rX} that the majority of sources lie in the
red AGN region, demonstrating the wider range of SED shapes and, in particular,
the large population of red AGN in this hard X-ray selected sample.
Many of the extended, optically bright sources which were below the AGN
region in Figure~\ref{fg:rX}, lie within the red AGN SED region.
This provides further evidence that they are predominantly 
obscured AGN for which the optical colors are brightened by host galaxy and/or
polarized light contributions (Kuraszkiewicz et al. 2009b, Smith et al. 2003).

Moving to longer wavelength bands, the X-ray fainter sources are preferentially
lost as the relatively higher flux limits cut down on the range of
observable IR fluxes.  This is most noticeable at 24$\mu$m
(Figure~\ref{fg:24X}) where the left-hand plot shows the 24$\mu$m 
vs the broad band X-ray flux and less than half the sample (333) are detected
in the IR. The labelling is the same as for Figure~\ref{fg:IRvsX_HR}.

\subsection{X-ray Sources Extended in Optical}

Of the 775 X-ray sources in the SWIRE/\chandra\ field, 107 were flagged as optically
extended by the source extraction
software (CASU pipeline Irwin \& Lewis 2001).
Visual inspection of the optical identifications of all the X-ray
sources led to inclusion of an additional 9 optically extended
sources, and exclusion
of 22 sources for false positives
arising from three causes: spikes from nearby saturated stars; multiple
closely spaced sources detected as a single, extended source; and proximity
to a different, unrelated extended source.  A further 2 sources were excluded as
it was determined that they are background AGN coincident with the extended
region of foreground galaxies.  The verified list thus consists of
92 optically extended sources.


In order to determine some of the general properties of the sample, we derive
luminosities from spectroscopic redshifts where available (29 of 92
sources, Polletta et al., in preparation)
and from photometric redshifts for all but 2 of the remaining sources,
which were too faint to obtain accurate photometry.  12 of these 92 sources
were identified as candidate non-nuclear ultraluminous X-ray sources (ULXs)
based upon their positions within the host galaxies.  9 of the 12 have
luminosities in excess of $\rm 10^{41}\ erg\ s^{-1}$ and are thus likely
not individual X-ray binaries.  A further 5 sources have luminosities
below $\rm 10^{41}\ erg\ s^{-1}$ and are thus also considered ULX candidates,
for a total of 8.  These sources will be discussed in more detail in
an upcoming paper (Kilgard et al., in preparation).


Most of the optically extended sources are at the optically
bright edge or below the AGN region, where galaxies are expected to lie.
Several of those within the AGN range are Sy2 galaxies.
We summarize the properties of two unusual sources
here.


CXOSW J104335.7+585249 
has an SED which fits a Sy 1.8 template (Polletta et al. 2006)
with a redshift, z$\sim 0.5$.
In the optical it is a bright, complex galaxy.
Its X-ray flux is quite bright (F(0.3-8keV)=3.3$\times 10^{-15}$ erg
cm$^{-2}$ s$^{-1}$) with a hardness ratio $\sim 0$
implying moderate absorption, \nh $\sim 10^{22}$ cm$^{-2}$, lower than
is usual for Type 1.8$-$2 AGN.


CXOSW J104552.4+590036 
is faint and somewhat fuzzy in the optical and fits to a
reddened AGN, MKN231-like SED (Polletta et al. 2006) at estimated redshift,
z$\sim 2.8$. Its X-ray hardness ratio, HR$\sim -0.4$ implies little
absorption. It has two nearby sources at 3.6 and 4.5 $\mu$m which may
effect these flux estimates.


\subsection{Radio-loudness}
\label{sec:rl}

It is common practice to divide AGN into two classes based on
the relative strength of their radio emission. 
However, there remain major questions concerning the reality of this
dichotomy, its cause and even the observational classification of radio-loud
vs radio-quiet AGN. For the former, physical properties such as larger
central black hole mass  (Lacy et al. 2001, Boroson 2002) and/or more rapid
black hole spin in radio-loud sources 
have been suggested to cause either a dichotomy or a progression between
one class and the other. 

From an observational standpoint, radio-loudness is a useful
tool to select out those sources with the strongest radio emission,
either having large, extended structures or core-dominated and beamed.
There are two alternative methods
for defining a radio-loud source. First using the radio-loudness, R$_L$,
traditionally measured as log of the the ratio of 6 cm radio
to optical flux (B magnitude, Wilkes \& Elvis 1987,
Kellermann \etal\ 1989), where
values $> 10$ indicate a radio-loud source. Second using the absolute radio
power where the number is around log L$_{20cm} \sim 31.6$ erg s$^{-1}$ Hz$^{-1}$
({\it e.g.} Zamfir \etal\ 2008). Since we do not have good redshift
estimates for the full sample, we will use the former definition here.

The bi-modality of the
distribution of radio-loudness in AGN is more/less pronounced
depending on the sample selection (Kellermann \etal\ 1989,
White \etal\ 2000, Lacy et al. 2001, Komossa \etal\ 2006,
Zamfir \etal\ 2008).
In addition, the use of the optical magnitude is questionable
given that it is strongly effected by redenning and/or orientation
(White \etal\ 2007) which are important constituents in IR/X-ray-selected
samples (Kuraszkiewicz \etal\ 2009a).
Figure~\ref{fg:hist_rl} shows the
distribution of radio-loudness (R$_L$= log[F$_{20cm}$/g$'$]) for the radio-detected
subset of the current X-ray selected
sample. The use
of g$'$ in place of B magnitude has a negligible effect on the parameter
given the small change in effective wavelength (4770\ang\ cf 4400\ang, Fukujita
\etal\ 1996). The use of 20cm rather than 6 cm radio flux is corrected
by assuming a spectral slope of 0.5 (where f$_{\nu}\propto \nu^{-\alpha}$) in
the radio. This shifts the standard division between radio-loud (RL) and
radio-quiet (RQ) from R$_L$=10 to R$_L$=19.
Within our radio-detected subset, 74 sources have no detection at
optical g$'$ but all of these are classifed as radio-loud based on 
the optical upper limit. The presence of
beaming in the radio and/or the relative strength of the
big blue bump, which is likely related to the accretion properties
and/or Eddington ratio, effect radio-loudness so that a
simple classification scheme does not ensure physically distinct sets of
sources (Falcke et al. 1996). In addition, this X-ray selected sample
includes a larger number of sources whose optical flux is faint in
comparison to optically-selected, Type 1 AGN.
To allow for the possible mis-classification of sources around the traditional
boundary, we divide the radio-quiet sources into an intermediate
class (RI): 1.9$<$R$_L <$19; and radio-quiet (RQ): R$_L <$ 1.0, as
labelled in Figure~\ref{fg:hist_rl},
again following Kellermann \etal\ (1989) but noting that the division
point is abritrary.
The R$_L$ distribution for this sample is not bi-modal
so there is no obvious division between radio-loud and radio-quiet
sources. The distribution appears asymmetric, has a large radio-loud fraction 
using this traditional classification, and has a long tail towards
radio-quiet sources.  Of the 
251 radio-detected X-ray sources in the
VLA field, 174 are radio-loud, 60 radio-intermediate and 17 radio-quiet.
Conservatively assuming that the non-radio-detected X-ray sources (
317) are radio-quiet,\footnote{We note that this is unlikely.
The upper limits for more than half these sources allow for them to be
radio-loud. Since the radio flux limit is a strong function of
position in the VLA field, those
towards the field edge are even more likely to be radio-loud.
}
$>$41\% of this
sample is classified as radio-loud using the traditional definition.
This is much higher than the $\sim 5-10$\% of radio-loud,
broad-lined AGN typically found in optical surveys (e.g. Peterson 1997).
Few of the radio-loud objects are strong radio sources, such as would be
found in radio catalogs. There are very few in the R$_L$ range of the
radio-quiet AGN in {\it e.g.} the optically selected PG AGN
(Kellermann \etal\ 1989).

\subsubsection{Radio and X-ray properties}
In Figure~\ref{fg:rx_rl} (left) we repeat Figure~\ref{fg:rX} but with the radio
classes indicated. The radio-loud sources lie preferentially
within the AGN region, those sources with low optical and
X-ray flux are almost exclusively radio-loud.
This is partially due to the fact that all radio-detected
sources with fluxes close to the optical and/or X-ray flux limits 
will be classified as radio-loud because 
the transition between radio-loud and radio-quiet classes 
falls below the faintest radio flux limit. Those which are radio-undetected
cannot be classified.

However, we must also examine the meaning of radio-loud in this
case. As discussed in Section~\ref{sec:rX},
the red and green lines in Figure~\ref{fg:rx_rl}
indicate the regions in which AGN with red SEDs and those with SEDs
typical of optically-selected, Type 1
AGN (respectively) lie as a function of redshift. The sources at the optical
and X-ray faint end lie beyond the range of optically-selected, Type 1
AGN SEDs (green line) but
within the range of the red SEDs (red line). We deduce that they are 
relatively optically faint (as shown in Kuraszkiewicz
et al. 2009a).  
As noted above, if the optical emission is relatively faint 
while the radio is uneffected,
the radio-loudness parameter, \rl , breaks down
as an indicator of bright radio emission. Thus an abnormally
large fraction of the sample are incorrectly classified as radio-loud,
as is clear in Figure~\ref{fg:rx_rl}.
While \rl~works well for optically-selected, Type 1, blue AGN, an
alternative radio-loudness indicator is needed for the remainder of
the AGN population.

\subsubsection{Radio-loudness determined using 24$\mu$m flux, \q24 .}
\label{sec:rlq24}

Given the limitations of \rl ,
it has been suggested that IR fluxes provide
a more stable measure of the radio-loudness. In particular
\q24 ~(=log(F$_{24\mu m}$/F$_{20cm}$), Appleton \etal\ 2004,
Kuraszkiewicz \etal\ 2009c) is minimally effected by Eddington ratio, 
host galaxy contribution, redshift and/or reddening.
The stability of the 24$\mu$m flux to obscuration is shown in 
Figure~\ref{fg:24XHR} where the ratio of the 24$\mu$m flux to the
broad (left) and hard (right) band X-ray fluxes is plotted
as a function of X-ray
hardness ratio. The broad band X-ray flux shows that hard, low-redshift
sources have higher 24$\mu$m flux relative to the X-ray while the
hard band X-ray flux does not. The effect is most likely due to
absorption in the X-ray band and indicates that the 24$\mu$m flux
is not effected by the X-ray absorbing material.

In Figure~\ref{fg:q24_rl} we show
the relation between \q24 ~and \rl ~for the 147 sources in our sample
for which data is available in all bands. The majority
of the sources are distributed around \q24 $\sim$ 1, consistent with the
IR-radio correlation for galaxies, on which radio-quiet AGN also lie
(Appleton et al. 2004). Kuraszkiewicz et al. (2009c) make a systematic
comparison of the two parameters for a large sample of blue, Type 1
quasars selected from the WSRT radio survey.
The historically-used transition from
radio-loud to radio-quiet, R$_L > 19$ for the current dataset,
and the equivalent value for
\q24 (=0.24$\pm$0.12) derived by  Kuraszkiewicz et al. (2009c) are indicated
by dotted lines. This figure dramatically shows the large
number of cross-over sources, those in the top right-hand quadrant, which are
radio-loud according to \rl ~and radio-quiet according to \q24 .
The conclusion that radio classification using \rl ~is incorrect for
these AGN is reinforced by the fact that they have 
\q24 $\sim 1$, as expected for radio-quiet AGN and galaxies (Appleton
et al. 2004). 
Using \q24 , 13 of the 147 sources 
are radio-loud, in excellent agreement
with the expected 5$-$10\% of radio-loud AGN based on 
optically-selected, Type 1 (unobscured, blue) AGN (Peterson 1997).
Figure~\ref{fg:rx_rl} (RHS) shows the same plot as the LHS but using
\q24 ~to define radio-loudness. A much smaller fraction of sources is
radio-loud and they are not systematically placed in the figure.

Figure~\ref{fg:IRvsX_rl} shows the {\it Spitzer} IRAC fluxes as a function of
X-ray flux (as in Figure~\ref{fg:IRvsX_HR}) with \q24 -classified, radio-loud
sources indicated. There is no visible relation between the radio
class and the relative IR and X-ray fluxes.

{\bf We conclude that \q24 ~is far superior to \rl ~for classification
of radio-loud vs radio-quiet AGN.}

\section{Conclusions}
We present a list of 775 \chandra\ X-ray sources in the SWIRE/\chandra\
medium-depth, X-ray survey. Cross-correlation with {\it Spitzer}, optical and
radio images of part/all of the field resulted in 771 (99\%) identifications in
at least one optical/IR band, 767 have IR counterparts visible in the
{\it Spitzer} data in at least one (IRAC) band and 
333 have 24$\mu$m with MIPS detections. Four of the sources have no
optical/IR counterpart down to our flux limits and 4 have an optical but no
IR couterpart. We present multi-wavelength flux measurements for 744
X-ray sources, all those which are uncontaminated, unconfused and above the
formal survey thresholds. The near-IR$-$X-ray datasets are
well-matched in flux limit and go deep into the AGN population,
providing an excellent dataset for multi-wavelength studies of the full
AGN population, which will be reported in a companion paper (Polletta et
al., in prep.). As in earlier surveys (DK04), there is no 
correlation between X-ray hardness and hard X-ray
flux in this sample,
confirming that hardness is predominantly caused by obscuration in the
X-rays. 

The very deep (2.7 $\mu$Jy at the field center) VLA data, covering
part of the \chandra\ field (Figure~\ref{fg:image}), 
results in 251 ($> 4 \sigma$) radio detections, $44$\% 
of the 568 X-ray sources in the VLA field.
Even the very deepest radio data cannot detect all the 
X-ray sources.
We demonstrate that the traditional radio-to-optical flux ratio, \rl,
used to define radio-loudness in AGN breaks down for a large proportion
of the sources in this X-ray selected sample due to the weakness of the optical
emission. Use of the 24$\mu$m flux in place of the optical, the
\q24 ~parameter (Appleton et al. 2004), brings the radio-loud
fraction down to expected levels (9\%) and is strongly preferred as an
indicator of radio-loudness for the full AGN population.

The wide range of optical/IR/X-ray SEDs in this X-ray selected sample is
demonstrated by the lack of any correlation between the optical or IR
flux and the X-ray flux. Comparison with tracks for
optically-selected, Type 1, and red AGN SEDs demonstrates
the predominance of red AGN in
this sample. There is a continuous distribution rather than
distinct classes of AGN, and the IR vs X-ray plots allow better discrimination
than the optical as the effects of obscuration are much lower.
Comparison of the source properties (X-ray
hardness, flux ratios and radio loudness) with the predictions
of the standard median SEDs allows us to broadly classify the source
types as a function of position in flux-flux plots.
A correlation between the 24$\mu$m flux and the X-ray hardness disappears
when the hard X-ray flux is used, demonstrating that it is due to
absorption in the X-rays while the 24$\mu$m flux remains uneffected. 
This reinforces the use of \q24 ~to define radio-loudness
due to its stability. 

\section*{Acknowledgements}
This work is
based in part on observations made with the Spitzer Space Telescope,
which is operated by the Jet Propulsion Laboratory, California
Institute of Technology under a contract with NASA.
We gratefully acknowledge the financial support of NASA Contract:
NAS8-03060 (\chandra\ X-ray Center, BW) and 
\chandra~GO grant: GO4-5158A (RK).
MP acknowledges financial contribution from contracts ASI-INAF I/016/07/0
and ASI-INAF  I/088/06/0.

\bibliography{refs}
\bibliographystyle{apj}

\begin{center}
\begin{deluxetable}{llllcc}
\tablecaption{X-ray Observations of the SWIRE/\chandra\ Fields \label{tab:swireobs}}
\tablewidth{0pt}
\tablehead{
  \colhead{ObsID} & \colhead{Aimpoint} & \colhead{Date Obs.} &
  \colhead{Exp. time} & \colhead{Galactic $N_H$} & \colhead{Roll Angle} \\
  \colhead{} & \colhead{J2000.0} & \colhead{UT} & \colhead{sec} & \colhead{($10^{19}\
cm^{-2}$)} &  }
\startdata
5023 & 10:46:02.15 +59:00:58.42 & 2004-09-12, 21:31:59 & 67146.56 & 6.6
& 17.60 \\
5024 & 10:44:48.27 +58:41:53.88 & 2004-09-16, 06:54:50 & 66297.89 & 6.4
& 17.60 \\
5025 & 10:46:41.57 +58:46:49.66 & 2004-09-17, 20:31:07 & 69257.73 & 6.3
& 17.60 \\
5026 & 10:48:34.91 +58:51:45.76 & 2004-09-18, 16:18:16 & 68863.14 & 6.4
& 17.60 \\
5027 & 10:44:08.82 +58:56:04.42 & 2004-09-20, 14:41:38 & 67076.31 & 6.7
& 20.07 \\
5028 & 10:47:55.60 +59:05:56.46 & 2004-09-23, 03:37:16 & 71145.48 & 6.5
& 21.20 \\
5029 & 10:43:29.40 +59:10:14.55 & 2004-09-24, 03:44:19 & 71127.09 & 7.1
& 21.20 \\
5030 & 10:45:22.74 +59:15:10.64 & 2004-09-25, 19:48:12 & 65748.49 & 6.9
& 21.20 \\
5031 & 10:47:16.03 +59:20:06.44 & 2004-09-26, 14:48:04 & 65393.51 & 6.8
& 21.20 \\
\enddata
\end{deluxetable}
\end{center}

\newpage
\begin{center}
\begin{deluxetable}{lr}
\tablecaption{Definition of Energy Bands and X-ray Colors
\label{tab:bands}}
\tablewidth{0pt}
\tabletypesize{\normalsize}
\tablecolumns{2}\tablehead{
\colhead{Band}&
\colhead{Definition}
}
\startdata
Broad (B)                     &    $0.3-8$ keV \\
Soft (S)                      &    $0.3-2.5$ keV \\
Hard (H)                      &    $2.5-8$ keV \\
Soft1 (S$_{1}$)               &    $0.3-0.9$ keV \\
Soft2 (S$_{2}$)               &    $0.9-2.5$ keV \\
\hline
Conventional Broad (Bc)       &    $0.5-8$ keV \\
Conventional Soft (Sc)        &    $0.5-2$ keV \\
Conventional Hard (Hc)        &    $2-8$ keV \\
\hline
Hardness Ratio $HR$           &    (Hc$-$Sc)$/$(Hc$+$Sc) \\
\enddata
\end{deluxetable}
\end{center}

\newpage
\include{table3_stub}

\include{table4_stub}

\newpage

\begin{deluxetable}{llll}
\small
\tablecaption{Power-law fits for SWIRE X-ray sources.  Errors quoted are 99\% confidence intervals computed using the \textit{project} function in
\it{Sherpa}. \label{tab:xspec}}
\tablewidth{0pc}
\tablehead{\colhead{Source} & \colhead{$N_H (10^{20} cm^{-2})$} & \colhead{$\Gamma$} & \colhead{$\chi^{2}$ / DOF} }
\startdata
CXOSW J104655.5+590301 
& $11.9^{+17.3}_{-11.9}$ & $1.65^{+0.38}_{-0.33}$ & 35.77/37 \\
CXOSW J104540.1+584254
$^1$ & $0.59^{+3.26}_{-0.0\tablenotemark{5}}$ & $2.27^{+0.25}_{-0.22}$ & 59.80/50 \\
CXOSW J104929.2+585338 
$^3$ & $0.59^{+6.55}_{-0.0\tablenotemark{5}}$ & $2.26^{+0.30}_{-0.14}$ & 89.79/70 \\
CXOSW J104803.4+585547 
$^2$ & $4.94^{+25.5}_{-0.0\tablenotemark{5}}$ & $2.04^{+0.81}_{-0.38}$ & 25.97/16 \\
CXOSW J104415.8+590101 
& $0.59^{+17.9}_{-0.0\tablenotemark{5}}$ & $1.78^{+0.49}_{-0.21}$ & 24.71/26 \\
CXOSW J104404.0+590241
$^4$ & $9.16^{+23.0}_{-9.16}$ & $2.10^{+0.62}_{-0.40}$ & 19.58/25 \\
CXOSW J104343.2+585535 
& $10.6^{+19.7}_{-10.6}$ & $1.75^{+0.48}_{-0.39}$ & 24.99/24 \\
CXOSW J104847.0+590455 
& $0.59^{+15.8}_{-0.0\tablenotemark{5}}$ & $2.26^{+0.54}_{-0.30}$ & 20.17/18 \\
CXOSW J104811.7+591046 
$^3$ & $0.59^{+16.0}_{-0.0\tablenotemark{5}}$ & $2.09^{+0.63}_{-0.25}$ & 26.74/19 \\
CXOSW J104422.6+591304 
& $13.0^{+18.5}_{-13.0}$ & $1.73^{+0.38}_{-0.32}$ & 53.39/65 \\
CXOSW J104321.3+590943 
& $2.81^{+9.29}_{-2.22\tablenotemark{5}}$ & $1.53^{+0.23}_{-0.16}$ & 67.08/83 \\
CXOSW J104503.4+591242 
$^3$ & $9.82^{+20.8}_{-9.82}$ & $1.81^{+0.47}_{-0.36}$ & 37.06/25 \\
CXOSW J104511.2+591625 
& $7.85^{+13.0}_{-7.85}$ & $1.91^{+0.38}_{-0.31}$ & 35.02/40\\
CXOSW J104743.5+591849 
& $0.59^{+14.1}_{-0.0\tablenotemark{5}}$ & $1.90^{+0.45}_{-0.21}$ & 15.49/23 \\
CXOSW J104713.6+591501 
& $0.59^{+9.43}_{-0.0\tablenotemark{5}}$ & $2.00^{+0.46}_{-0.32}$ & 19.44/17 \\
\enddata
\begin{itemize}
\item 1. 
Fit with pl+nei ($\Gamma$ = 2.13, kT = 0.065, reduced $\chi^2$ = 0.87), pl+bbody ($\Gamma$ = 2.66, kT = 2.77, reduced $\chi^2$ = 0.97), or 2 power-laws ($\Gamma_1$ = 5.00, $\Gamma_2$ = 1.89, reduced $\chi^2$ = 0.85).
\item 2. 
Fit by a MEKAL model with $kT = 3.15 keV$.
\item 3. No good fit was found 
\item 4. 
Shows evidence for a spectral line at 1.14 keV.  It is not required for the fit,
but is preferred at around the 1.5 $\sigma$ level.
\item 5. Hard parameter space minimum
\end{itemize}
\end{deluxetable}

\newpage
\begin{center}
\begin{deluxetable}{cccrrrr}
\tablecaption{List of the Best Fit Parameters \label{tab:lnls}}
\tablewidth{0pt}
\tabletypesize{\large}
\tablehead{
\colhead{DATA}&
\colhead{$\Gamma_{ph}$}&
\colhead{Band$^1$}&
\colhead{$K^2$}& 
\colhead{$\gamma_{1}^3$}&
\colhead{$\gamma_{2}^4$}&
\colhead{$S_{b}^5$}\\ 
}
\startdata
SWIRE & 1.4 &       S &1011$^{+  88}_{ -86}$ & 1.89$^{+0.08}_{-0.07}$ &         2.35$^{+0.47}_{-0.30}$ & 12.7$^{+11.5}_{-5.3}$ \\ 
      &     &      H &1263$^{+ 481}_{-351}$ & 1.65$^{+0.21}_{-0.20}$ &         2.49$^{+0.24}_{-0.16}$ &  7.5$^{+ 2.5}_{-0.7}$\\ 
      &     &      B & 1540$^{+ 170}_{-351}$ & 1.67$^{+0.06}_{-0.07}$ &         2.71$^{+0.52}_{-0.53}$ & 23.4$^{+ 5.2}_{-24.9}$\\ 
\hline
      & 1.7 &       S &  974$^{+  86}_{ -85}$ & 1.85$^{+0.07}_{-0.07}$ &    2.39$^{+0.46}_{-0.32}$ & 13.5$^{+ 9.2}_{-5.6}$ \\ 
      &     &   H & 1762$^{+ 604}_{-448}$ & 1.88$^{+0.20}_{-0.20}$ &    2.47$^{+0.39}_{-0.19}$ &  7.9$^{+ 5.1}_{-1.4}$\\ 
      &     &       B & 1489$^{+ 161}_{-448}$ & 1.68$^{+0.06}_{-0.06}$ &    2.70$^{+0.52}_{-0.41}$ & 22.2$^{+ 4.8}_{-24.1}$\\ 
\hline
      & 1.4   &       Sc & 560$^{+  45}_{ -45}$ & 1.65$^{+0.09}_{-0.14}$ &         2.42$^{+0.49}_{-0.34}$ &  7.1$^{+ 2.9}_{-3.2}$ \\ 
      & &       Hc &1722$^{+ 322}_{-632}$ & 1.84$^{+0.10}_{-0.31}$ &         2.81$^{+0.72}_{-0.54}$ & 14.9$^{+ 7.5}_{-7.6}$\\ 
      &    &       Bc & 1548$^{+ 247}_{-632}$ & 1.66$^{+0.09}_{-0.13}$ &         2.40$^{+0.25}_{-0.23}$ & 14.8$^{+ 6.2}_{-4.7}$\\ 
\hline
      & 1.7 &       Sc &  579$^{+  48}_{ -48}$ & 1.66$^{+0.09}_{-0.14}$ &    2.41$^{+0.49}_{-0.33}$ &  7.3$^{+ 3.1}_{-3.3}$ \\ 
      &     &       Hc & 1384$^{+  69}_{ -83}$ & 1.72$^{+0.03}_{-0.05}$ &    2.59$^{+0.26}_{-0.23}$ &  9.5$^{+ 1.9}_{-3.0}$\\ 
      &    &       Bc & 1384$^{+ 160}_{ -83}$ & 1.66$^{+0.08}_{-0.10}$ &    2.41$^{+0.35}_{-0.23}$ & 13.4$^{+ 6.1}_{-4.1}$\\ 
\hline
\hline
ChaMP$^6$ & 1.4 &       S & 769$^{+  14}_{ -14}$ & 1.57$^{+0.01}_{-0.01}$ &         2.41$^{+0.05}_{-0.05}$ &  9.9$^{+ 0.7}_{-1.6}$ \\ 
      &     &      H &1828$^{+  48}_{ -43}$ & 1.81$^{+0.01}_{-0.01}$ &         2.58$^{+0.05}_{-0.05}$ & 14.2$^{+ 0.9}_{-1.1}$\\ 
      &     &      B & 1614$^{+  28}_{ -43}$ & 1.65$^{+0.01}_{-0.01}$ &         2.44$^{+0.06}_{-0.05}$ & 25.0$^{+ 1.9}_{-1.9}$\\ 
\hline
      & 1.7 &       S &  783$^{+  15}_{ -15}$ & 1.58$^{+0.01}_{-0.01}$ &    2.42$^{+0.05}_{-0.05}$ & 10.5$^{+ 0.8}_{-0.8}$ \\ 
      &     &   H & 1774$^{+  44}_{ -41}$ & 1.80$^{+0.01}_{-0.01}$ &    2.58$^{+0.05}_{-0.05}$ & 13.5$^{+ 0.9}_{-0.9}$\\ 
      &     &       B & 1505$^{+  25}_{ -41}$ & 1.65$^{+0.01}_{-0.01}$ &    2.45$^{+0.06}_{-0.05}$ & 21.9$^{+ 1.7}_{-1.7}$\\ 
\hline
      & 1.4   &       Sc & 607$^{+  12}_{ -12}$ & 1.54$^{+0.02}_{-0.02}$ &         2.36$^{+0.05}_{-0.05}$ &  6.8$^{+ 0.5}_{-0.5}$ \\ 
      & &       Hc &2040$^{+  50}_{ -50}$ & 1.82$^{+0.01}_{-0.01}$ &         2.65$^{+0.07}_{-0.07}$ & 19.2$^{+ 6.3}_{-1.8}$\\ 
      &    &       Bc & 1557$^{+  28}_{ -50}$ & 1.64$^{+0.01}_{-0.01}$ &         2.48$^{+0.05}_{-0.05}$ & 22.9$^{+ 1.6}_{-1.6}$\\ 
\hline
      & 1.7 &       Sc &  612$^{+  12}_{ -12}$ & 1.53$^{+0.02}_{-0.02}$ &    2.36$^{+0.05}_{-0.04}$ &  6.7$^{+ 0.5}_{-0.5}$ \\ 
      &     &       Hc & 1932$^{+  46}_{ -48}$ & 1.82$^{+0.01}_{-0.01}$ &    2.64$^{+0.07}_{-0.07}$ & 17.8$^{+ 4.4}_{-1.7}$\\ 
      &    &       Bc & 1407$^{+  25}_{ -48}$ & 1.64$^{+0.01}_{-0.01}$ &    2.48$^{+0.05}_{-0.05}$ & 19.2$^{+ 1.3}_{-1.4}$\\ 
\hline
ChaMP+ & 1.4 &       Sc & 571$^{+  11}_{ -11}$ & 1.49$^{+0.02}_{-0.02}$ &
  2.36$^{+0.05}_{-0.05}$ &  6.5$^{+ 0.4}_{-0.4}$ \\
CDFs$^1$   &     &       Hc &1258$^{+  29}_{ -29}$ & 1.58$^{+0.01}_{-0.01}$ &
  2.59$^{+0.06}_{-0.05}$ & 14.4$^{+ 0.9}_{-0.9}$\\
\hline
\enddata\\
\begin{minipage}{5.5in}
1: X-ray energy band (Table~\ref{tab:bands})\\
2: normalization constant\\
3: faint power index of a broken power law\\
4: bright power index of a broken power law\\
5: break flux in units of $10^{-15}$ $\rm erg~cm^{-2}~sec^{-1}$\\
6. from Kim et al., 2007b
\end{minipage}
\end{deluxetable}
\end{center}

\newpage
\include{table7_stub}

\clearpage

\clearpage

\clearpage

\clearpage
\clearpage

\clearpage
\begin{figure}
\epsscale{0.8}
\plotone{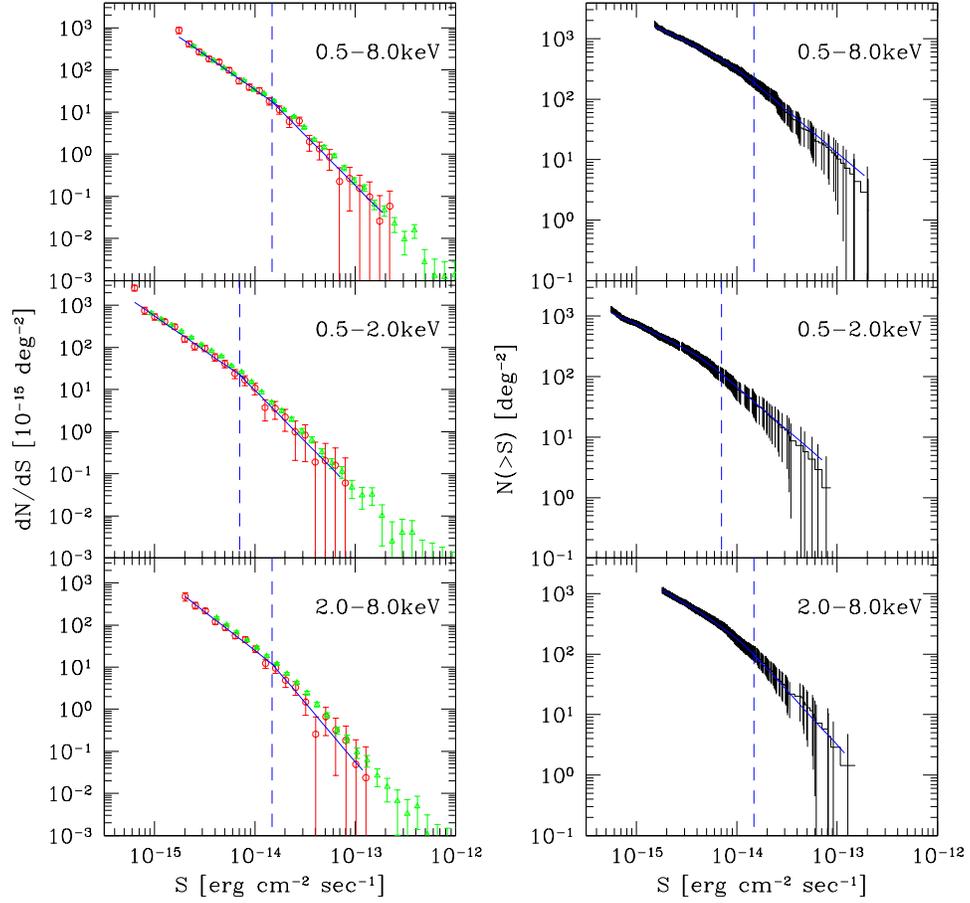}
\caption{The differential (left) and cumulative (right) logN-logS
  distribution for the
SWIRE/\chandra\ field (red) in comparison with those from the ChaMP full
catalog (green, from Kim \etal\ 2007a). The blue dashed line indicates
the best fit break energy. The blue solid line indicates the best fit
broken power law.
}
\label{fg:lnls}
\end{figure}

\clearpage
\begin{figure}
\epsscale{1.0}
\plotone{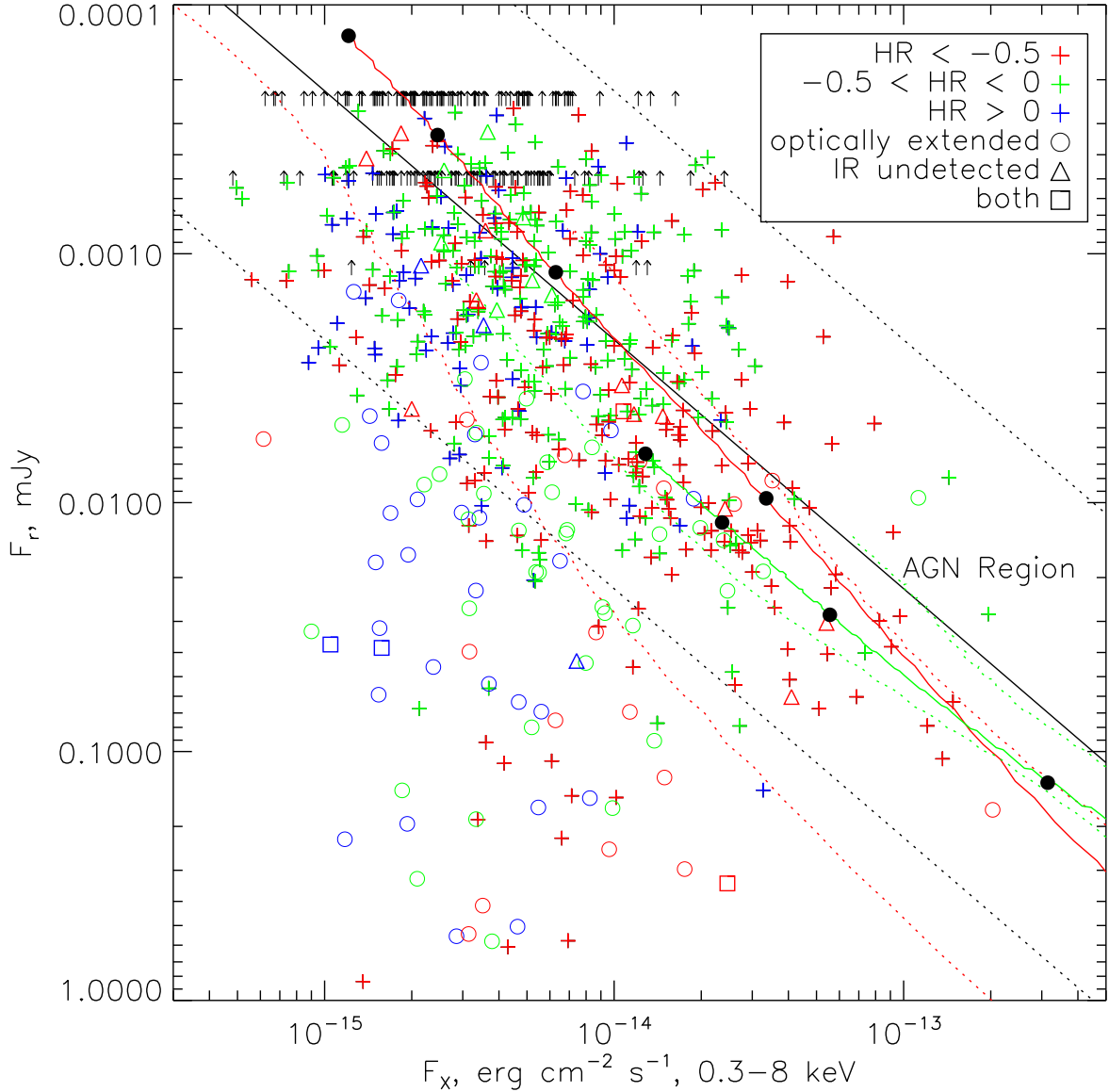}
\caption{Optical r$'$-band as a function of broad band X-ray flux 
for the SWIRE X-ray sources. 
The black solid and dotted lines show
f$_x$/f$_{r}$=0.1,1,10, providing a guide to where
extragalactic sources (mostly AGN) 
generally lie in this figure (``AGN region", Stocke et al. 1991). 
Optically extended and IR undetected
sources are indicated by different symbols as listed, with squares
indicating those which are both.
X-ray hard (HR$>$0), intermediate ($-0.5<$HR$<0$)
and soft (HR$<-0.5$) sources are labelled by color.
Tracks for a median ($\pm 90$\%) AGN SED (Elvis \etal\ 1994)
and a median red ($\pm 90$\%)
AGN SED (Kuraskiewicz \etal\ 2009a) as a function of redshift (up to z=4)
are displayed as green and red solid (dotted) lines respectively (see
Section~\ref{sec:rX}). Black dots indicate redshifts of 1,2,3,4 from
the lower right.
Optical upper limits (indicating the 90\% completeness level)
are shown as black arrows with the different levels
indicating different
optical datasets.}
\label{fg:rX}
\end{figure}

\clearpage
\begin{figure}
\epsscale{1.0}
\plottwo{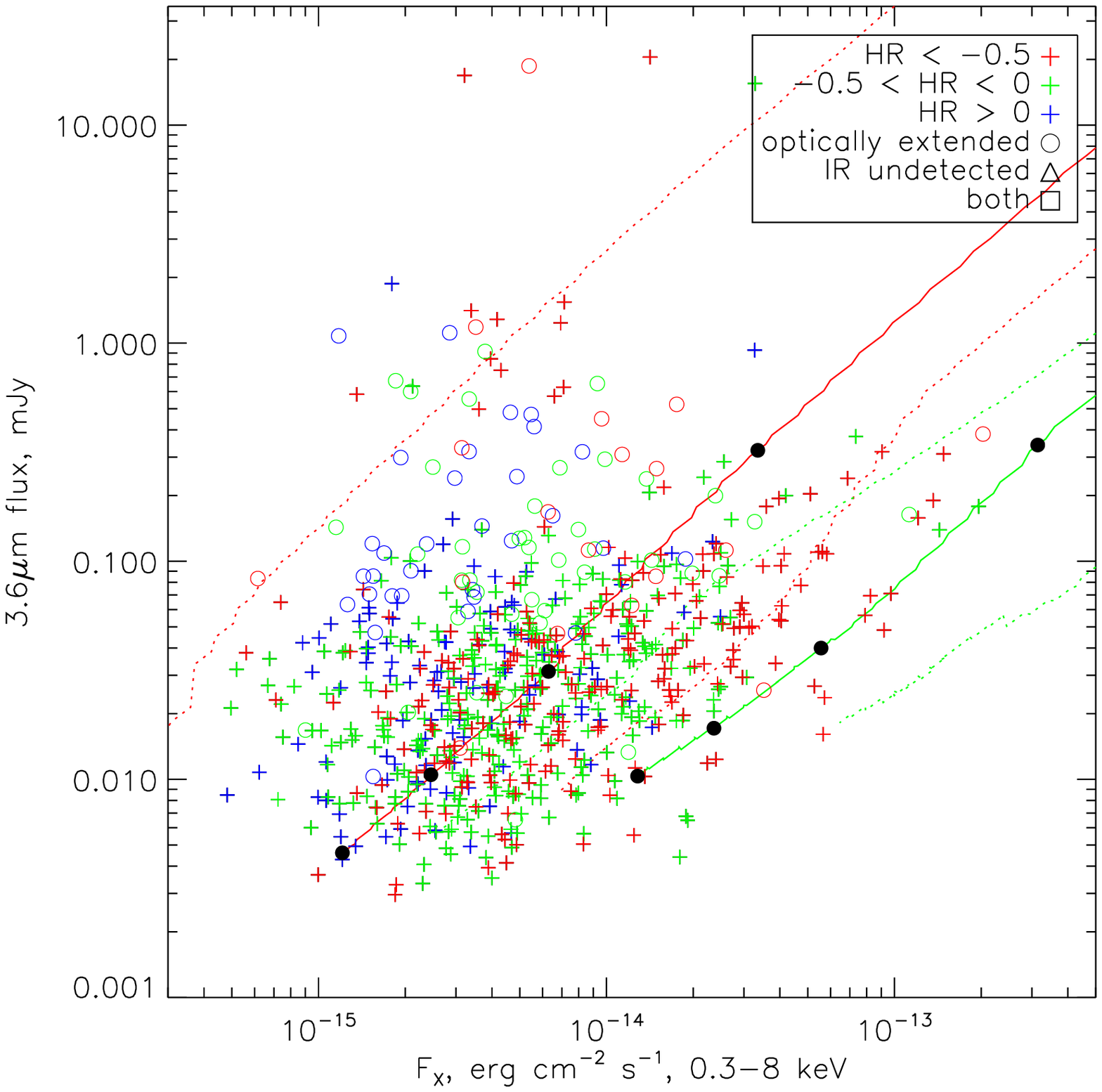}{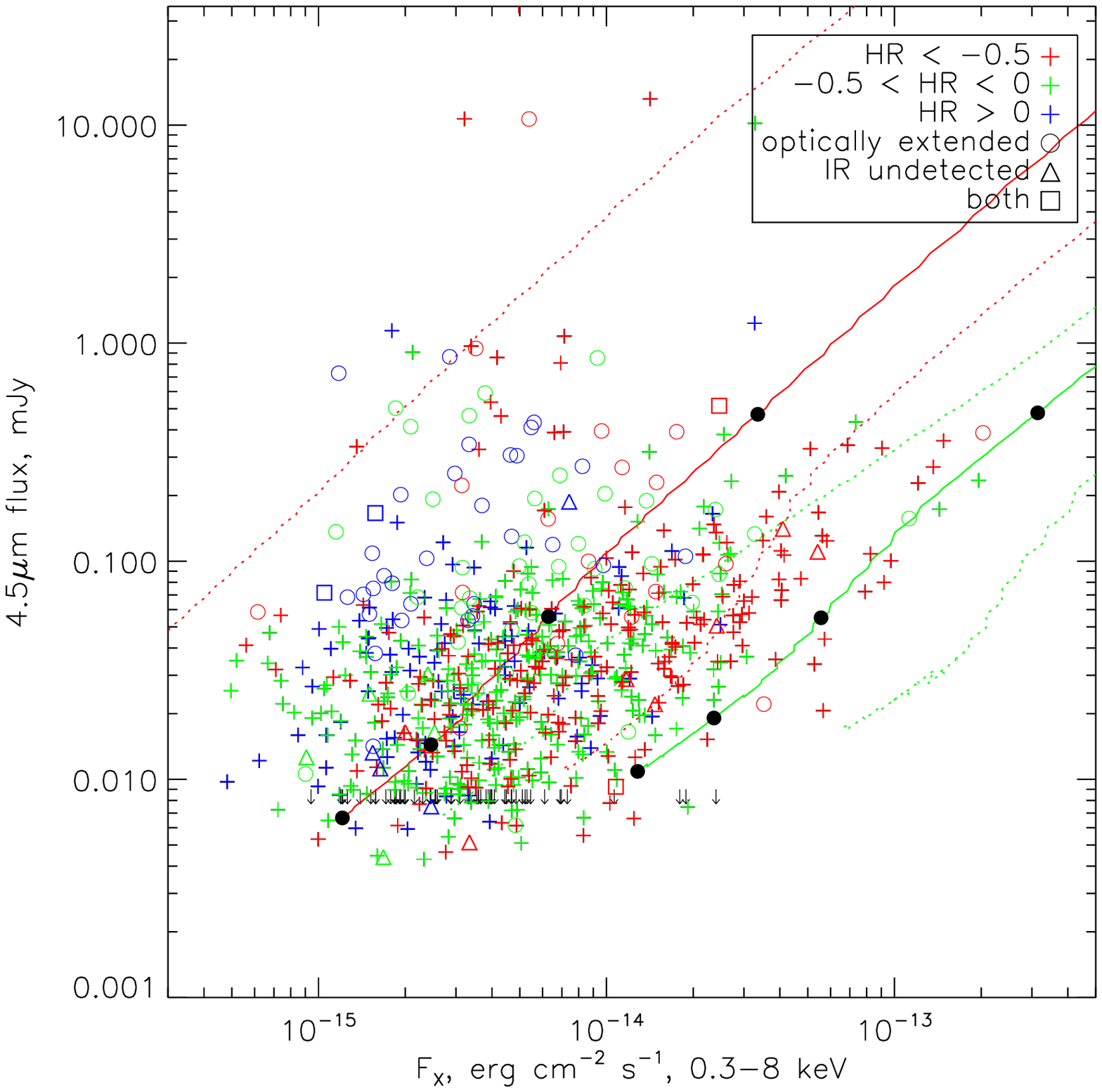}
\plottwo{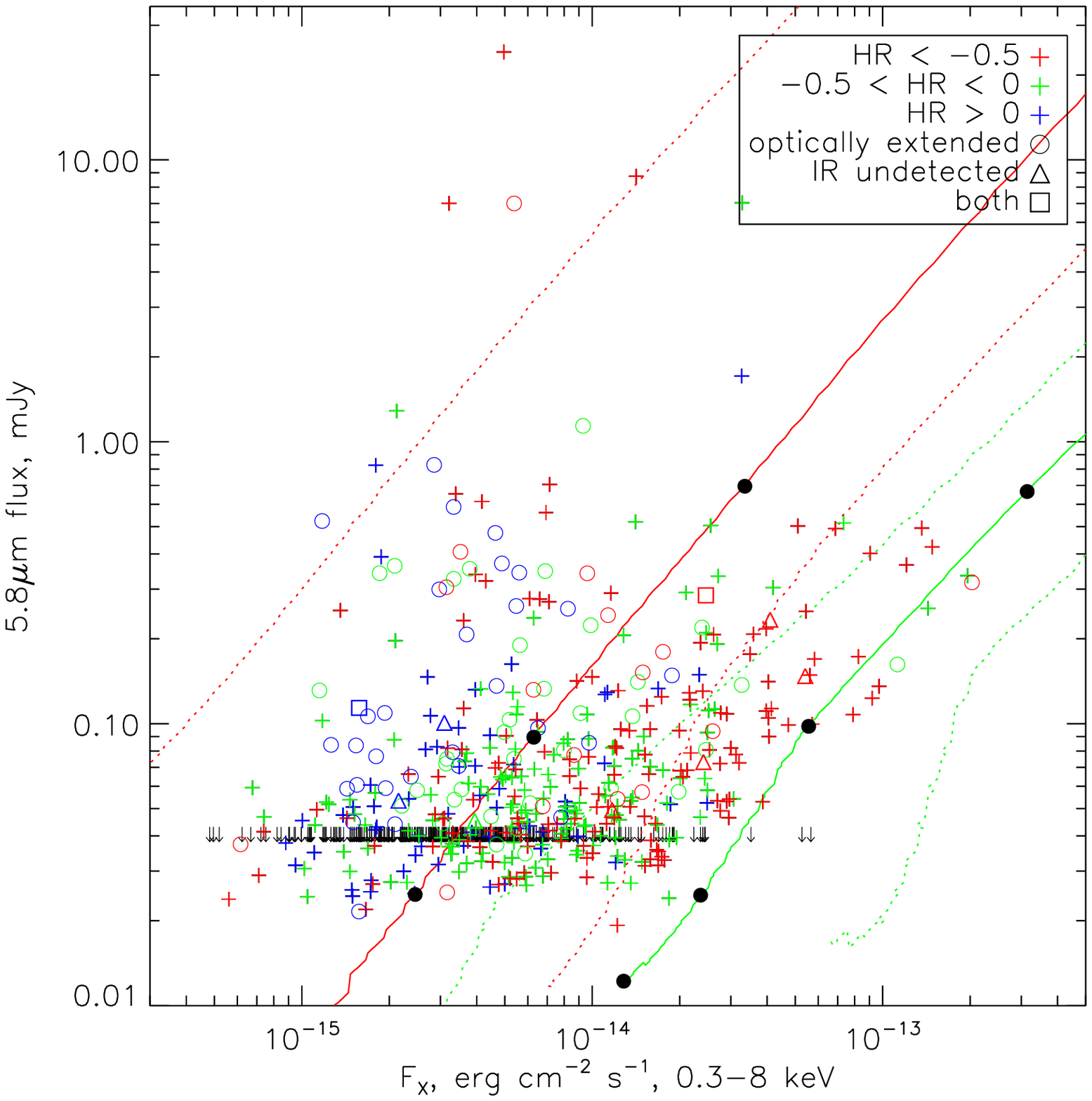}{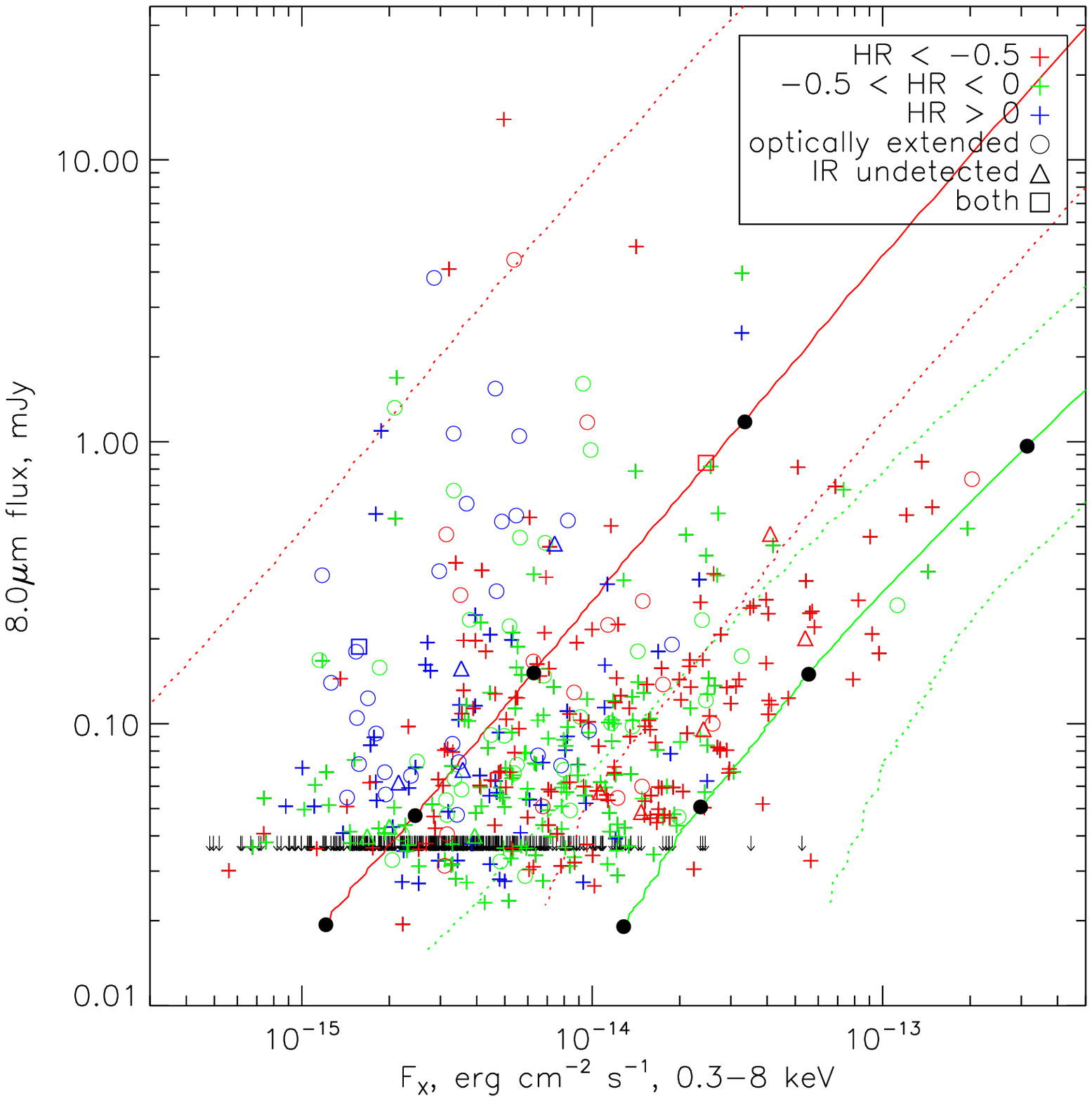}
\caption{IR as a function of
broad band X-ray flux for the SWIRE X-ray sources in each IRAC waveband.
Symbols and lines are as in Figure~\ref{fg:rX}.
Upper limits (5$\sigma$) are indicated by black arrows.
The tracks for optically-selected, type 1 and red AGN SEDs cover largely distinct regions in these figures.
Optically extended sources tend to be IR bright for their X-ray flux and
mostly lie within the range of the red AGN SEDs, implying the presence
of an AGN which is not visible in the optical.
}
\label{fg:IRvsX_HR}
\end{figure}

\clearpage
\begin{figure}
\epsscale{1.0}
\plottwo{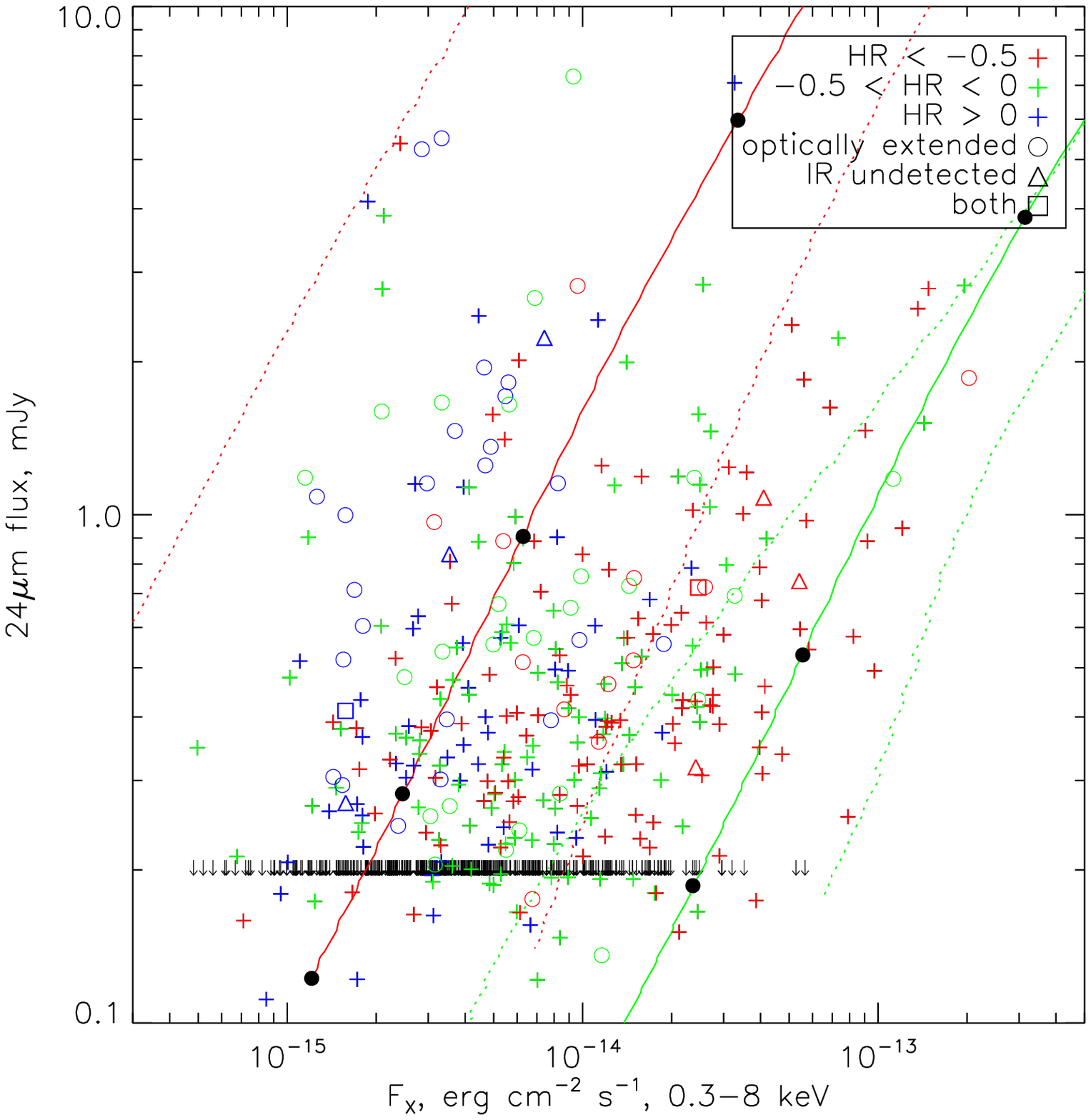}{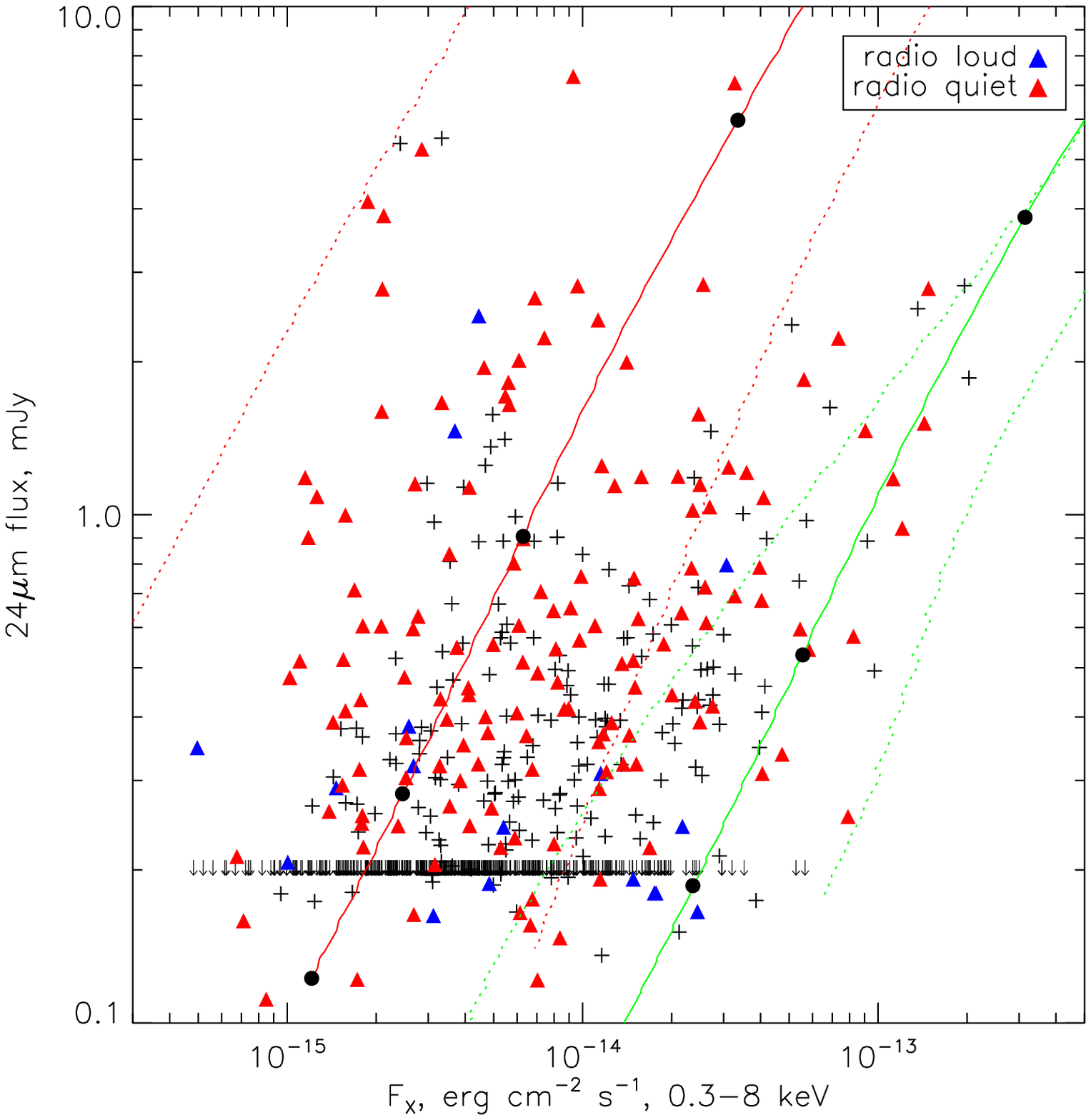}
\caption{IR 24$\mu$m as a function of 
broad band X-ray flux for the SWIRE X-ray sources.
Upper limits (5$\sigma$) are indicated by black arrows.
Left: Symbols and lines are as in Figure~\ref{fg:rX}; 
Right: Sources are labelled according to their
\q24 ~radio-loudness (Section~\ref{sec:rlq24}).}
\label{fg:24X}
\end{figure}

\clearpage
\begin{figure}
\epsscale{1.0}
\plotone{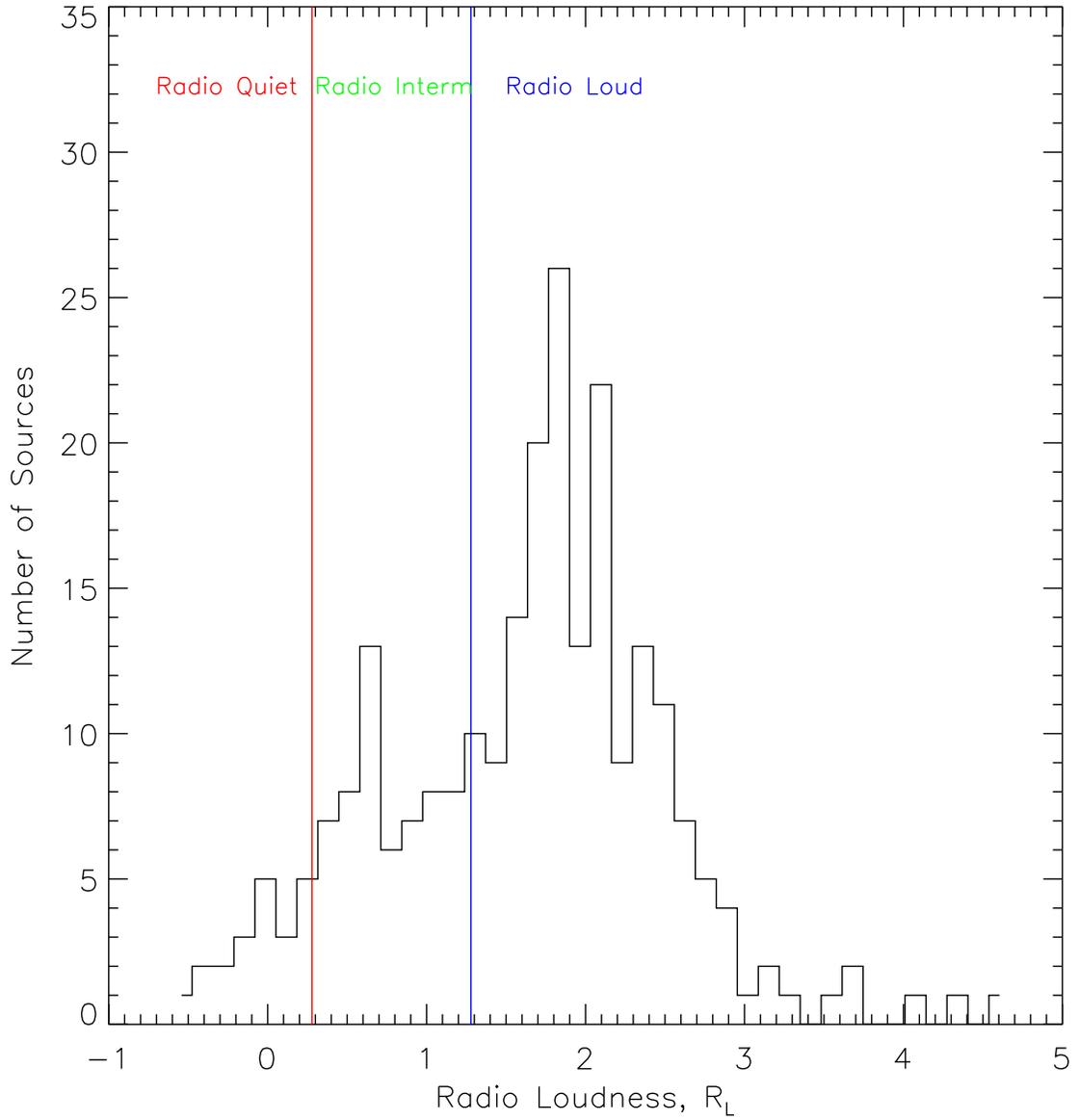}
\caption{Histogram of the traditional measure of radio-loudness (R$_L$ = log[f$_{20cm}$/f$_{g'}$])
for all
radio and optically detected sources in the sample. The divisions between
radio-loud (R$_L > 19$), radio-intermediate ($1.9 < $R$_L < 19$) and
radio-quiet (R$_L < 1.9$) for this dataset are
shown by vertical blue and red lines respectively (Section~\ref{sec:rl}).}
\label{fg:hist_rl}
\end{figure}

\clearpage
\begin{figure}
\epsscale{1.0}
\plottwo{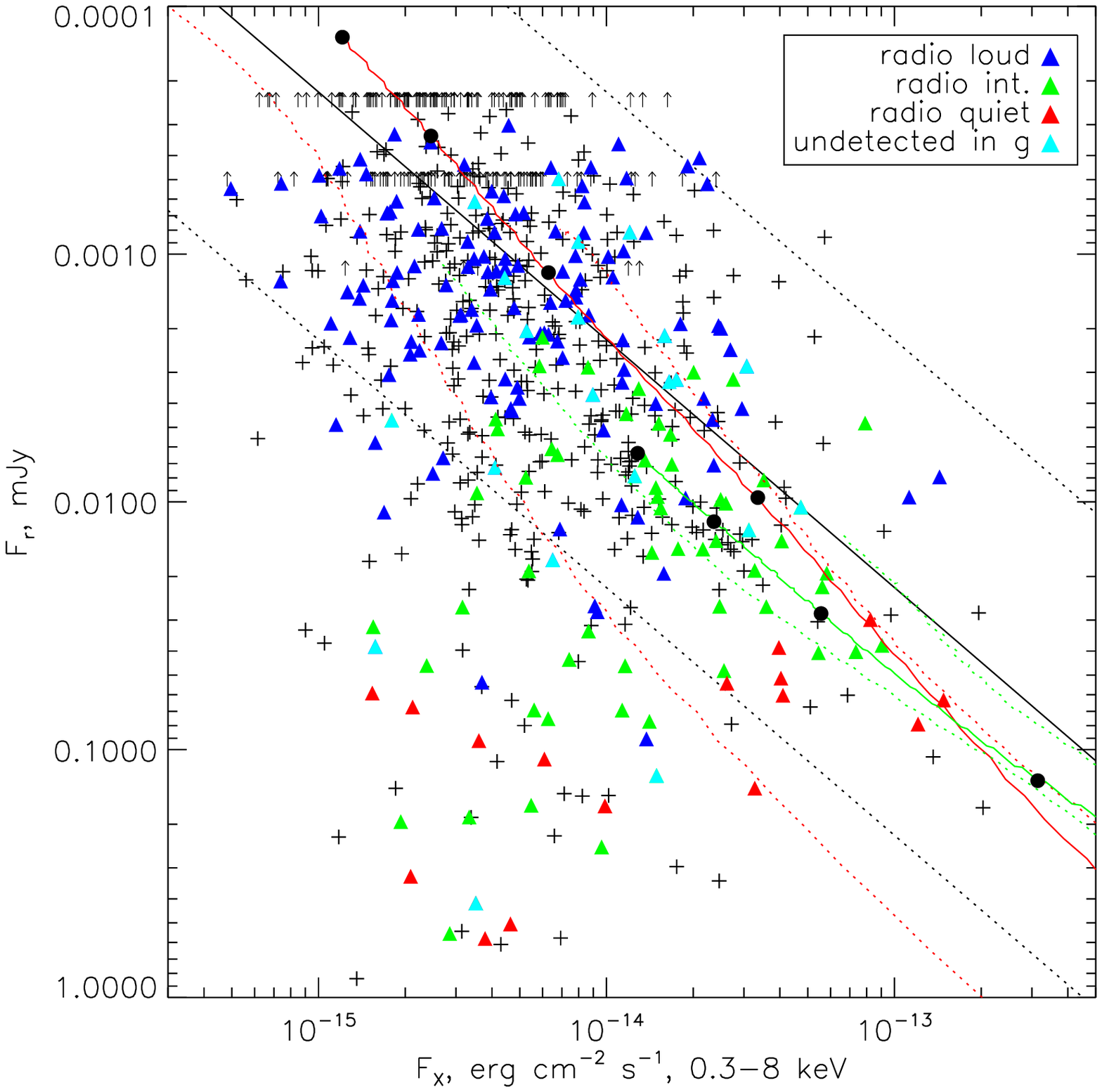}{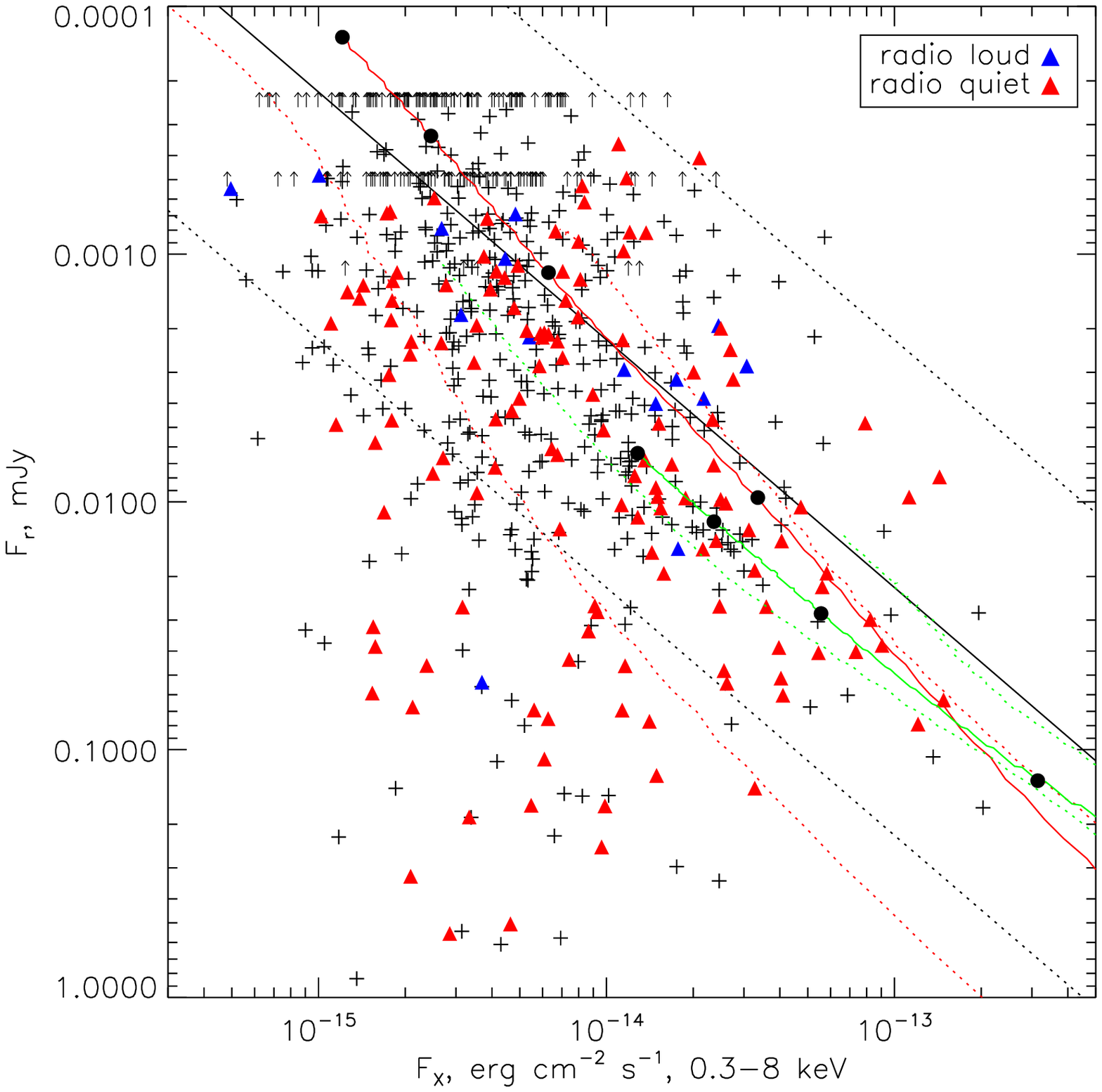}
\caption{Optical r$'$-band as a function of broad band X-ray flux 
for the SWIRE X-ray sources with sources labeled according to their
radio class: Left: using the traditional radio-loudness (\rl )
parameter (Figure~\ref{fg:hist_rl});
Right: using \q24 $> 0.24$, to define radio-loud AGN.
Other lines and symbols are as in Figure~\ref{fg:rX}.
The large number of radio loud sources as
defined by \rl ~(left), lying at faint optical fluxes
in the red AGN region, are mostly classified as radio-quiet
when the more stable \q24 
~classification method is used to define radio-loud (right).
}
\label{fg:rx_rl}
\end{figure}

\clearpage
\begin{figure}
\epsscale{1.0}
\plottwo{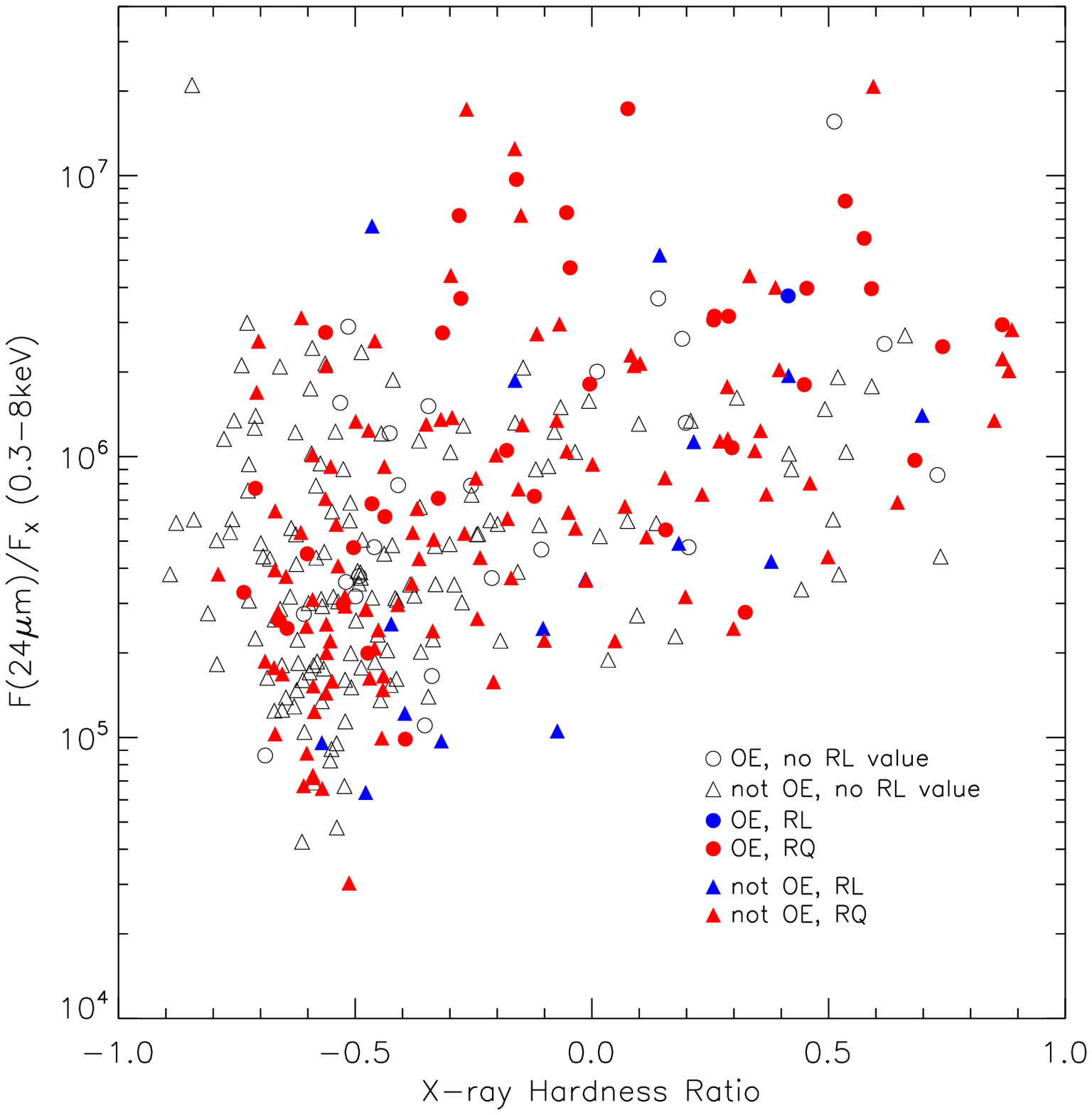}{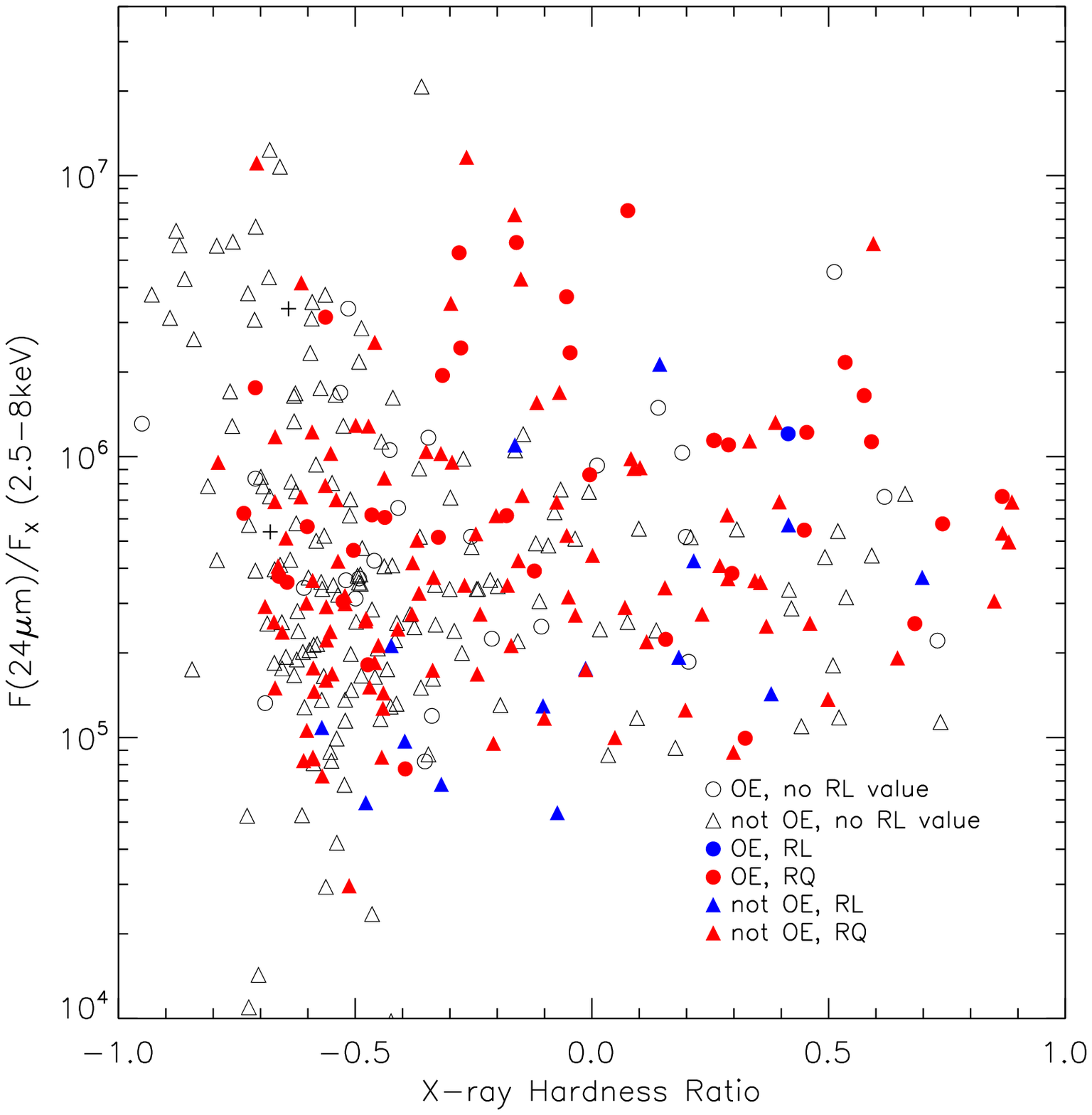}
\caption{The dimensionless ratio of 24$\mu$m to effective 
broad (left) and hard (right)
X-ray flux density (determined in $\mu$Jy at 1 keV from the hard and
broad-band fluxes respectively, assuming $\Gamma=1.7$) as a function of 
X-ray hardness ratio for those X-ray sources with
24$\mu$m identifications. 
The tendency for hard sources to have higher 24$\mu$m to X-ray
flux ratios is present only for broad-band X-ray flux, indicating
that this is an absorption effect in the X-ray alone.
The colors indicate the \q24 ~radio class
(Section~\ref{sec:rlq24}). Symbols are defined in the legend, where
OE=optically extended.
}
\label{fg:24XHR}
\end{figure}

\clearpage
\begin{figure}
\epsscale{1.0}
\plotone{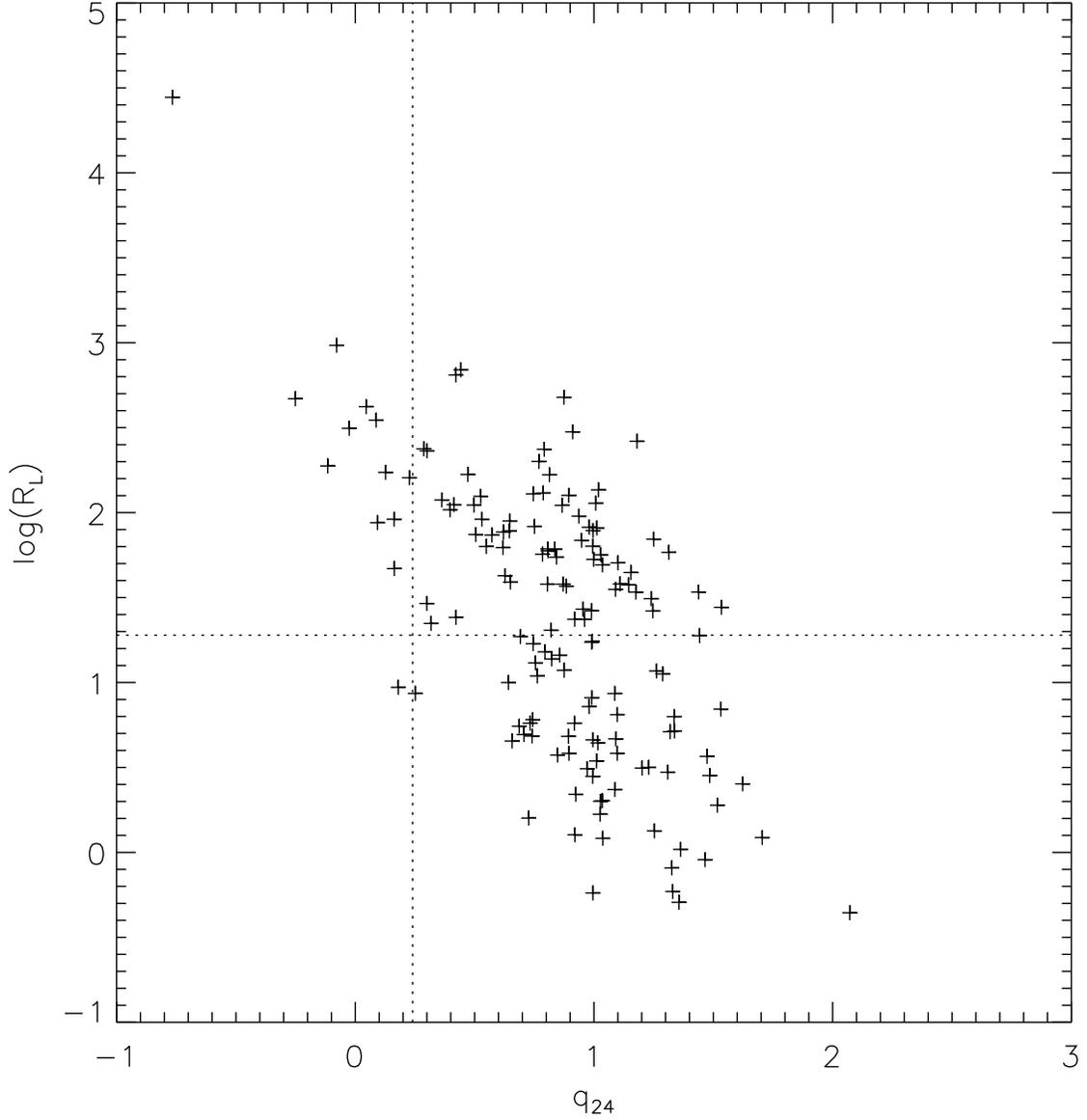}
\caption{Radio-loudness (\rl ) as a function of mid-IR-to-radio flux ratio (\q24 ).
The radio-loud/quiet
boundaries for each parameter (\rl ~= 19, \q24 ~= 0.24$\pm$0.12), the
latter based on a comparison using a large sample of type 1 quasars
(Kuraszkiewicz et al. 2009c),
are marked with dashed lines. The large number of
cross-over sources in our X-ray selected sample
are visible in the top right quadrant. }
\label{fg:q24_rl}
\end{figure}

\clearpage
\begin{figure}
\epsscale{1.0}
\plottwo{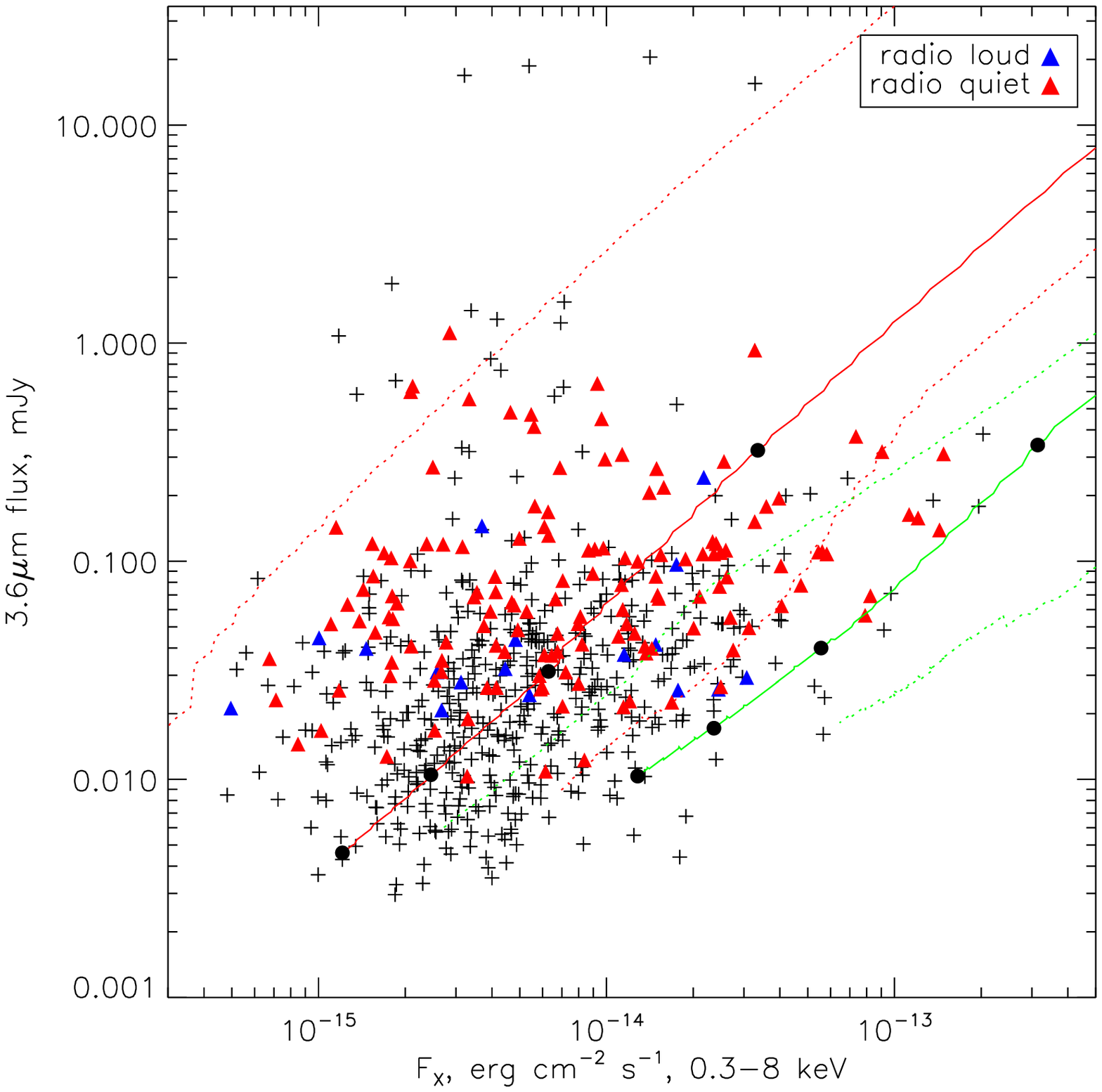}{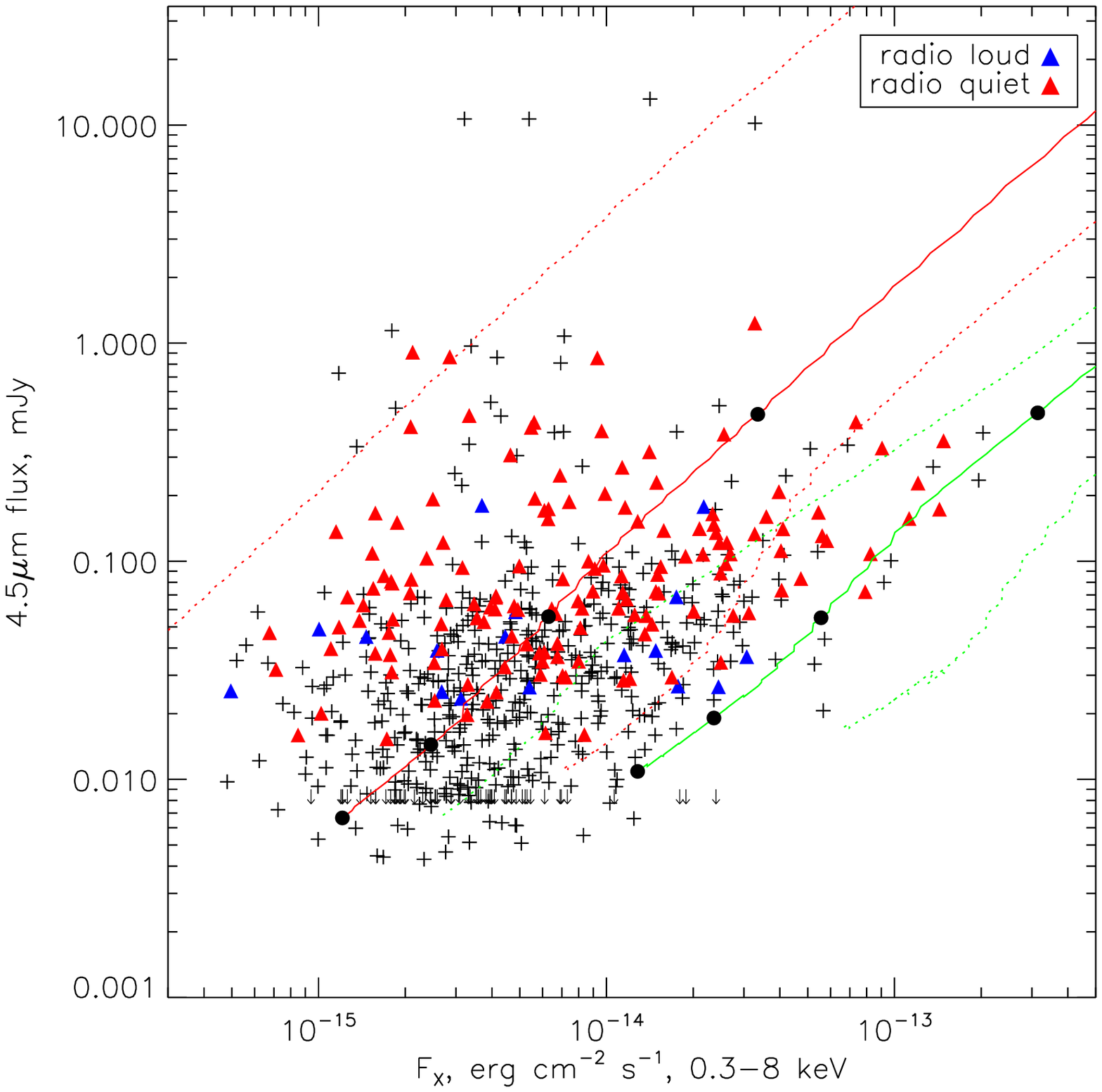}
\plottwo{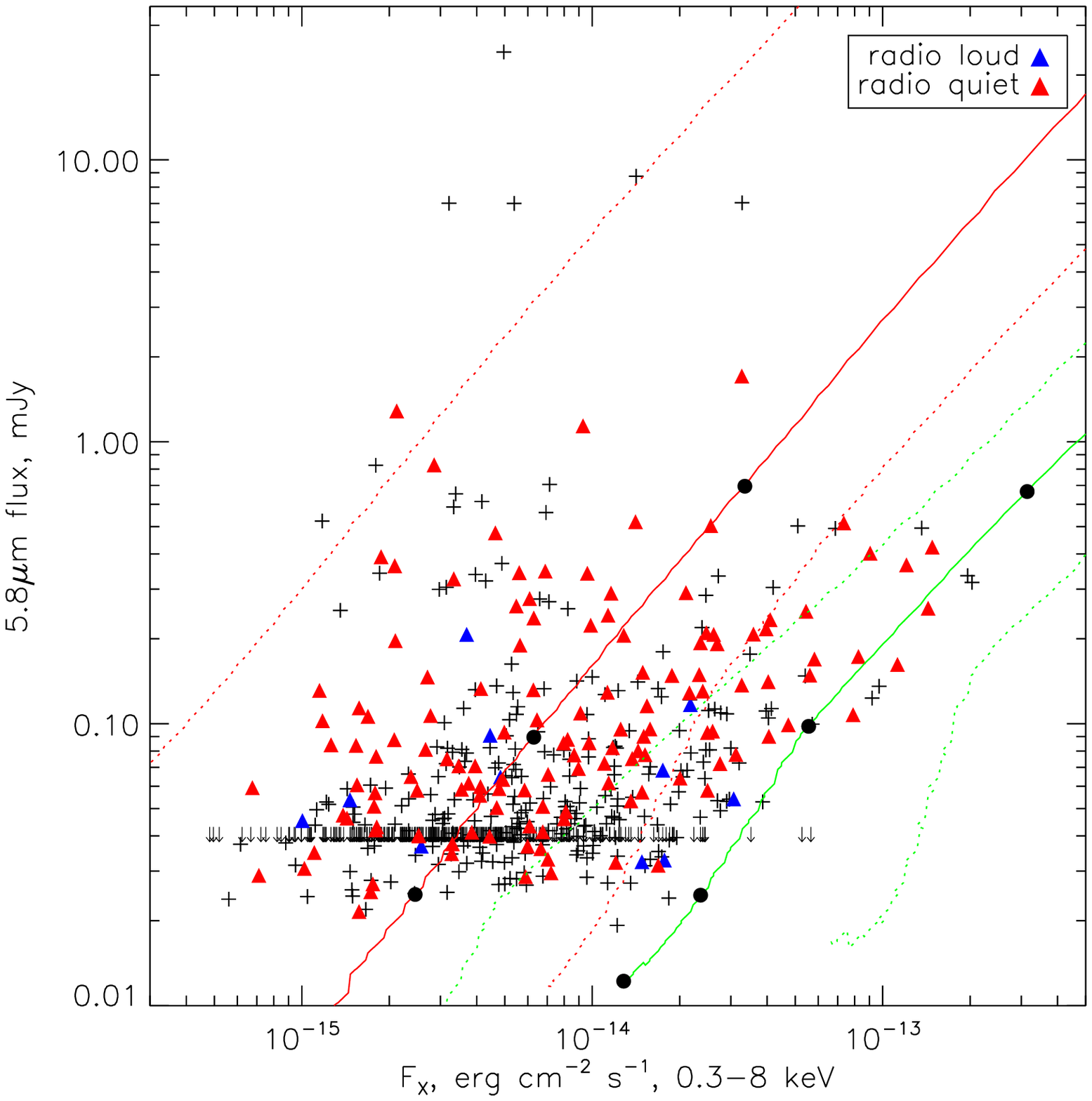}{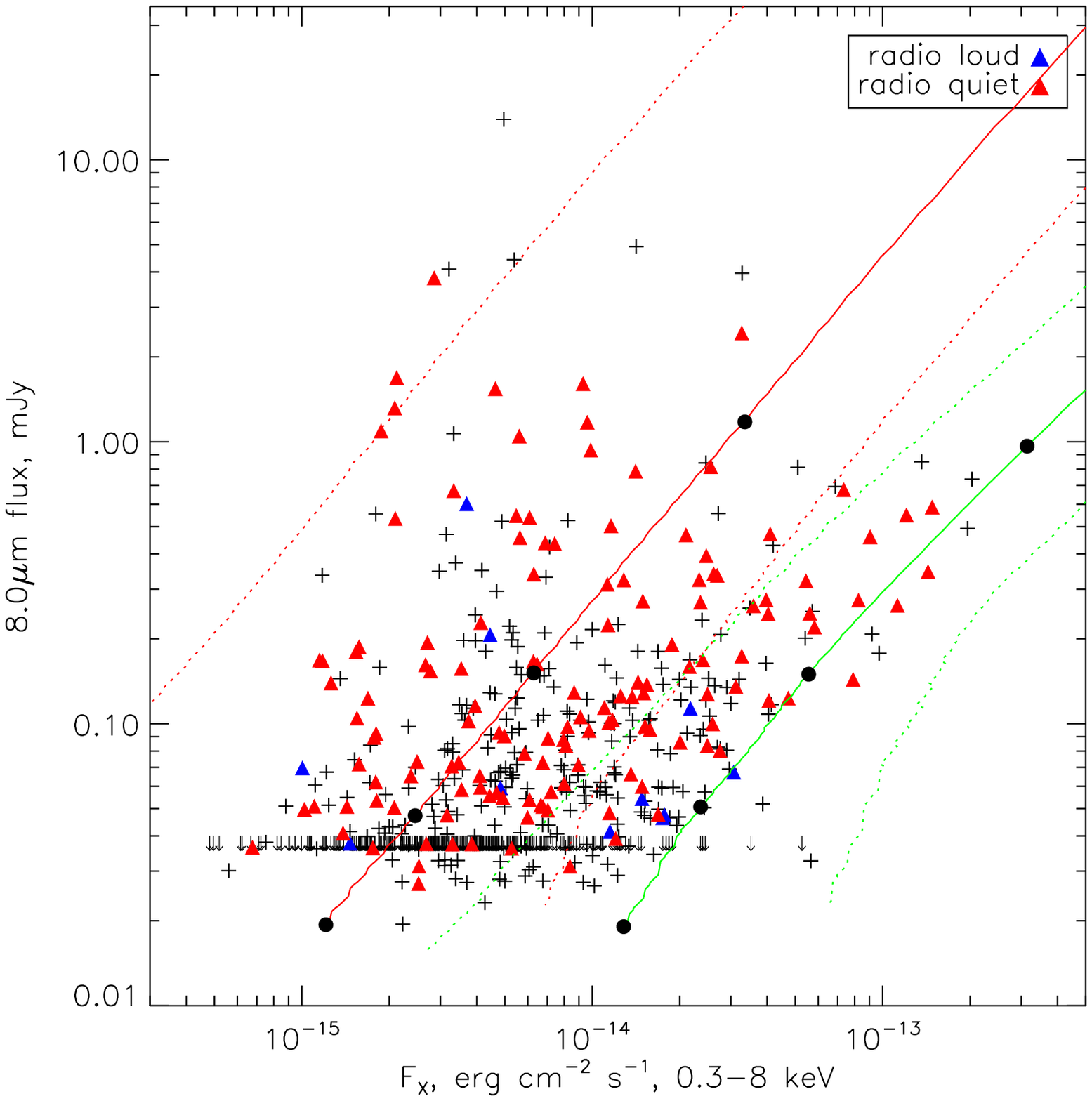}
\caption{Spitzer IR as a function of
broad band X-ray flux for the SWIRE X-ray sources in each IRAC
band. Symbols are
as in Figure~\ref{fg:IRvsX_HR}, but 
with radio class defined using \q24 . 
}
\label{fg:IRvsX_rl}
\end{figure}

\end{document}

%% file: table3_stub.tex
\begin{deluxetable}{llllll}
\tabletypesize{\scriptsize}
\tablecaption{SWIRE X-ray sources \label{tab:swirex}}
\tablewidth{0pt}
\tablehead{
  \colhead{Name} & \colhead{RA} & \colhead{Dec} & \colhead{Pos. err}$^1$ &
  \colhead{X-ray counts} & \colhead{Count rate} \\
  \colhead{} & \colhead{J2000} & \colhead{J2000} & \colhead{arcsec} & \colhead{0.3-8.0 keV} & \colhead{counts/sec} }
\startdata
CXOSWJ104629.3+585941 & 10:46:29.27 & 58:59:41.3 &  0.16 &  131.3 $\pm$ 12.6 & 1.96E-03 $\pm$ 1.88E-04 \\
CXOSWJ104616.4+585921 & 10:46:16.41 & 58:59:21.1 &  0.24 &   27.9 $\pm$  6.6 & 4.16E-04 $\pm$ 9.76E-05 \\
CXOSWJ104613.5+585941 & 10:46:13.47 & 58:59:41.6 &  0.09 &  150.6 $\pm$ 13.4 & 2.24E-03 $\pm$ 2.00E-04 \\
CXOSWJ104647.0+585618 & 10:46:46.95 & 58:56:18.3 &  0.49 &  106.1 $\pm$ 12.0 & 1.58E-03 $\pm$ 1.79E-04 \\
CXOSWJ104638.5+585642 & 10:46:38.47 & 58:56:42.9 &  0.51 &   50.5 $\pm$  8.8 & 7.53E-04 $\pm$ 1.30E-04 \\
CXOSWJ104633.2+585815 & 10:46:33.18 & 58:58:15.9 &  0.26 &   91.4 $\pm$ 10.8 & 1.36E-03 $\pm$ 1.61E-04 \\
CXOSWJ104622.4+590052 & 10:46:22.40 & 59:00:52.7 &  0.26 &   25.9 $\pm$  6.3 & 3.85E-04 $\pm$ 9.35E-05 \\
CXOSWJ104622.0+585630 & 10:46:21.97 & 58:56:30.1 &  0.31 &   80.4 $\pm$ 10.3 & 1.20E-03 $\pm$ 1.53E-04 \\
CXOSWJ104658.9+585959 & 10:46:58.91 & 58:59:59.6 &  0.83 &   33.6 $\pm$  7.9 & 5.00E-04 $\pm$ 1.17E-04 \\
CXOSWJ104649.5+585532 & 10:46:49.49 & 58:55:32.2 &  0.86 &   45.1 $\pm$  9.2 & 6.71E-04 $\pm$ 1.37E-04 \\
CXOSWJ104647.3+590122 & 10:46:47.30 & 59:01:22.4 &  0.49 &   67.5 $\pm$  9.7 & 1.01E-03 $\pm$ 1.44E-04 \\
CXOSWJ104644.1+590028 & 10:46:44.07 & 59:00:28.3 &  0.49 &   33.1 $\pm$  7.2 & 4.94E-04 $\pm$ 1.07E-04 \\
CXOSWJ104637.5+585914 & 10:46:37.49 & 58:59:14.8 &  0.65 &   21.9 $\pm$  6.1 & 3.26E-04 $\pm$ 9.07E-05 \\
CXOSWJ104707.3+590014 & 10:47:07.27 & 59:00:14.6 &  0.85 &   49.8 $\pm$  9.5 & 7.42E-04 $\pm$ 1.41E-04 \\
CXOSWJ104654.6+585804 & 10:46:54.57 & 58:58:04.3 &  0.75 &   36.1 $\pm$  8.2 & 5.38E-04 $\pm$ 1.22E-04 \\
CXOSWJ104653.3+585735 & 10:46:53.32 & 58:57:35.2 &  0.92 &   21.0 $\pm$  7.2 & 3.13E-04 $\pm$ 1.07E-04 \\
CXOSWJ104651.0+585824 & 10:46:50.98 & 58:58:24.9 &  1.00 &   11.2 $\pm$  5.9 & 1.66E-04 $\pm$ 8.83E-05 \\
CXOSWJ104648.1+590103 & 10:46:48.12 & 59:01:03.0 &  1.08 &   16.6 $\pm$  6.0 & 2.47E-04 $\pm$ 8.95E-05 \\
CXOSWJ104705.3+585850 & 10:47:05.26 & 58:58:50.4 &  1.59 &   16.2 $\pm$  7.1 & 2.42E-04 $\pm$ 1.06E-04 \\
CXOSWJ104703.1+585622 & 10:47:03.05 & 58:56:22.3 &  1.72 &   20.3 $\pm$  8.0 & 3.02E-04 $\pm$ 1.19E-04 \\
CXOSWJ104658.4+585914 & 10:46:58.37 & 58:59:14.7 &  1.56 &   22.9 $\pm$  7.0 & 3.41E-04 $\pm$ 1.05E-04 \\
CXOSWJ104632.7+585409 & 10:46:32.71 & 58:54:09.4 &  1.63 &   26.5 $\pm$  7.8 & 3.94E-04 $\pm$ 1.15E-04 \\
CXOSWJ104611.1+585517 & 10:46:11.06 & 58:55:17.5 &  0.44 &   58.4 $\pm$  9.1 & 8.69E-04 $\pm$ 1.36E-04 \\
CXOSWJ104607.6+585602 & 10:46:07.60 & 58:56:02.8 &  0.63 &   15.6 $\pm$  5.5 & 2.33E-04 $\pm$ 8.13E-05 \\
CXOSWJ104557.0+590000 & 10:45:57.00 & 59:00:00.4 &  0.19 &   23.6 $\pm$  6.1 & 3.52E-04 $\pm$ 9.06E-05 \\
\enddata \\
\begin{minipage}{5.5in}
1: The position error was determined as described in MK07
\end{minipage}
\end{deluxetable}

%% file: table4_stub.tex
\begin{deluxetable}{llllcr}
\tabletypesize{\scriptsize}
\tablecaption{SWIRE X-ray sources \label{tab:flux}}
\tablewidth{0pt}
\tablehead{
  \colhead{Name} & \colhead{Flux Density$^1$} & \colhead{Broad-band Flux$^1$}
  & \colhead{Soft Flux$^1$} & \colhead{Hard Flux$^1$} & \colhead{Hardness} \\
  \colhead{} & \colhead{1 keV} & \colhead{0.3-8 keV} &
  \colhead{0.3-2.5 keV} & \colhead{2.5-8 keV} & \colhead{Ratio$^2$} }
\startdata
CXOSW J104629.3+585941 &   5.76E-15 &   2.24E-14 &   1.33E-14 &   6.45E-15 & $-0.70^{+0.05}_{-0.07}$ \\
CXOSW J104616.4+585921 &   1.23E-15 &   4.65E-15 &   2.92E-15 &   9.12E-16 & $-0.75^{+0.10}_{-0.16}$ \\
CXOSW J104613.5+585941 &   6.61E-15 &   2.49E-14 &   1.19E-14 &   1.56E-14 & $-0.37^{+0.07}_{-0.08}$ \\
CXOSW J104647.0+585618 &   4.65E-15 &   1.91E-14 &   1.00E-14 &   9.40E-15 & $-0.50^{+0.09}_{-0.10}$ \\
CXOSW J104638.5+585642 &   2.22E-15 &   8.93E-15 &   4.50E-15 &   4.90E-15 & $-0.43^{+0.13}_{-0.15}$ \\
CXOSW J104633.2+585815 &   4.01E-15 &   1.58E-14 &   8.67E-15 &   6.67E-15 & $-0.56^{+0.08}_{-0.10}$ \\
CXOSW J104622.4+590052 &   1.13E-15 &   4.69E-15 &   1.71E-15 &   4.44E-15 & $-0.04^{+0.19}_{-0.20}$ \\
CXOSW J104622.0+585630 &   3.53E-15 &   1.49E-14 &   7.88E-15 &   7.18E-15 & $-0.50^{+0.09}_{-0.11}$ \\
CXOSW J104658.9+585959 &   1.47E-15 &   6.14E-15 &   3.40E-15 &   2.50E-15 & $-0.56^{+0.17}_{-0.19}$ \\
CXOSW J104649.5+585532 &   1.98E-15 &   8.71E-15 &   5.24E-15 &   2.36E-15 & $-0.69^{+0.14}_{-0.17}$ \\
CXOSW J104647.3+590122 &   2.96E-15 &   1.20E-14 &   2.91E-15 &   1.56E-14 & $0.30^{+0.12}_{-0.12}$ \\
CXOSW J104644.1+590028 &   1.45E-15 &   5.84E-15 &   2.35E-15 &   4.90E-15 & $-0.15^{+0.17}_{-0.18}$ \\
CXOSW J104637.5+585914 &   9.60E-16 &   3.80E-15 &   1.68E-15 &   2.74E-15 & $-0.25^{+0.21}_{-0.23}$ \\
CXOSW J104707.3+590014 &   2.19E-15 &   9.31E-15 &   2.18E-15 &   1.23E-14 & $0.32^{+0.17}_{-0.15}$ \\
CXOSW J104654.6+585804 &   1.58E-15 &   6.83E-15 &   4.81E-15 &  -         & $-0.87^{+0.03}_{-0.13}$ \\
CXOSW J104653.3+585735 &   9.23E-16 &   3.89E-15 &   2.16E-15 &   1.59E-15 & $-0.52^{+0.22}_{-0.29}$ \\
CXOSW J104651.0+585824 &   4.90E-16 &   2.05E-15 &   5.92E-16 &   2.39E-15 & $0.14^{+0.40}_{-0.36}$ \\
CXOSW J104648.1+590103 &   7.29E-16 &   2.96E-15 &   1.31E-15 &   2.15E-15 & $-0.25^{+0.27}_{-0.28}$ \\
CXOSW J104705.3+585850 &   7.12E-16 &   3.06E-15 &   2.47E-15 &  -         & $-0.78^{+0.05}_{-0.22}$ \\
CXOSW J104703.1+585622 &   8.91E-16 &   3.85E-15 &   9.86E-16 &   4.83E-15 & $0.23^{+0.34}_{-0.29}$ \\
CXOSW J104658.4+585914 &   1.01E-15 &   4.19E-15 &   2.41E-15 &   1.43E-15 & $-0.59^{+0.19}_{-0.26}$ \\
CXOSW J104632.7+585409 &   1.16E-15 &   4.90E-15 &   1.89E-15 &   4.34E-15 & $-0.11^{+0.25}_{-0.24}$ \\
CXOSW J104611.1+585517 &   2.56E-15 &   1.01E-14 &   5.53E-15 &   4.36E-15 & $-0.55^{+0.11}_{-0.13}$ \\
CXOSW J104607.6+585602 &   6.85E-16 &   2.89E-15 &   1.22E-15 &   2.24E-15 & $-0.20^{+0.25}_{-0.27}$ \\
CXOSW J104557.0+590000 &   1.04E-15 &   7.83E-15 &   4.02E-15 &   4.11E-15 & $-0.44^{+0.16}_{-0.20}$ \\
\enddata \\
\begin{minipage}{5.5in}
1: Fluxes assume Galactic N$_{\rm H}$ and $\Gamma$ = 1.7, in units
of erg cm$^{-2}$ s$^{-1}$ (flux) and  erg cm$^{-2}$ s$^{-1}$ keV$^{-1}$
(flux density) \\
2: Hardness Ratio = (H-S)/(H+S), where H = counts in hard band
(2.5-8keV), S = Counts in the soft band (0.3-2.5keV), determined using
a Bayesian approach which is applicable for low count sources
(see text)
\end{minipage}
\end{deluxetable}

%% file: table7_stub.tex
\begin{deluxetable}{lllllllll}
\tabletypesize{\scriptsize}
\tablecaption{SWIRE X-ray sources: Multi-Wavelength Fluxes$^1$ \label{tab:swireID}}
\tablewidth{0pt}
\rotate
\tablehead{
  \colhead{Name} & \colhead{SWIRE ID} & \colhead{$\Delta$ pos.$^2$} & \colhead{r' mag} & \colhead{g' mag} & \colhead{3.6 $\mu$m flux} & \colhead{8 $\mu$m flux} & \colhead{24 $\mu$m flux} & \colhead{20cm flux} \\
  \colhead{} & \colhead{} & \colhead{arcsec} & \colhead{} & \colhead{} & \colhead{$\mu$Jy} & \colhead{$\mu$Jy} & \colhead{$\mu$Jy} & \colhead{$\mu$Jy} }
\
\startdata
CXOSW J104629.3+585941 & 561274 &  0.26 &      24.43 $\pm$  0.09 &      24.89 $\pm$  0.09 &      11.86 $\pm$  0.45 &      30.50 $\pm$  3.18 &     -209.0 $\pm$  0.00 &       13.6 $\pm$    3.4 \\
CXOSW J104616.4+585921 & 559904 &  0.87 &      22.16 $\pm$  0.02 &      24.00 $\pm$  0.07 &      15.33 $\pm$  0.53 &     -40.00 $\pm$  0.00 &     -209.0 $\pm$  0.00 &       39.5 $\pm$    7.6 \\
CXOSW J104613.5+585941 & 559874 &  0.25 &      22.97 $\pm$  0.03 &      23.34 $\pm$  0.06 &      26.57 $\pm$  0.66 &     127.34 $\pm$  3.61 &      1147. $\pm$  17.0 &       92.0 $\pm$    5.5 \\
CXOSW J104647.0+585618 & 560498 &  0.94 &      24.61 $\pm$  0.11 &      25.15 $\pm$  0.11 &       6.49 $\pm$  0.40 &      47.76 $\pm$  3.43 &     -209.0 $\pm$  0.00 &       23.0 $\pm$    4.2 \\
CXOSW J104638.5+585642 & 560041 &  0.44 &     -25.20 $\pm$  0.00 &     -25.90 $\pm$  0.00 &      25.96 $\pm$  0.63 &      65.64 $\pm$  3.08 &      193.6 $\pm$  17.3 &      -14.7 $\pm$    0.0 \\
CXOSW J104633.2+585815 & 560612 &  0.94 &      20.49 $\pm$  0.02 &      21.90 $\pm$  0.02 &     217.91 $\pm$  1.81 &      95.33 $\pm$  5.67 &      1188. $\pm$  19.3 &      266.0 $\pm$    8.2 \\
CXOSW J104622.4+590052 & 561464 &  0.50 &     -25.20 $\pm$  0.00 &     -25.90 $\pm$  0.00 &       4.87 $\pm$  0.50 &     -40.00 $\pm$  0.00 &     -209.0 $\pm$  0.00 &       15.7 $\pm$    3.9 \\
CXOSW J104622.0+585630 & 558585 &  0.31 &      18.45 $\pm$  0.02 &      19.76 $\pm$  0.06 &     265.98 $\pm$  1.85 &     272.25 $\pm$  5.80 &      750.9 $\pm$  18.1 &      421.2 $\pm$    3.3 \\
CXOSW J104658.9+585959 & 564107 &  0.88 &     -25.20 $\pm$  0.00 &     -25.90 $\pm$  0.00 &      10.88 $\pm$  0.44 &     -40.00 $\pm$  0.00 &      164.9 $\pm$  16.6 &       23.0 $\pm$    3.8 \\
CXOSW J104649.5+585532 & 560200 &  0.98 &      22.00 $\pm$  0.03 &      22.65 $\pm$  0.03 &      25.08 $\pm$  0.57 &      32.19 $\pm$  3.06 &     -209.0 $\pm$  0.00 &      -16.0 $\pm$    0.0 \\
CXOSW J104647.3+590122 & 564038 &  2.57 &      23.94 $\pm$  0.06 &      24.30 $\pm$  0.07 &      22.85 $\pm$  0.51 &      39.14 $\pm$  3.29 &      312.1 $\pm$  16.9 &       93.0 $\pm$    6.1 \\
CXOSW J104644.1+590028 & 563124 &  1.63 &      22.59 $\pm$  0.04 &      23.52 $\pm$  0.05 &      29.87 $\pm$  0.67 &      78.35 $\pm$  3.42 &      803.9 $\pm$  17.5 &       29.9 $\pm$    5.6 \\
CXOSW J104637.5+585914 & 561669 &  1.56 &      23.39 $\pm$  0.04 &      24.43 $\pm$  0.08 &      33.64 $\pm$  0.68 &     -40.00 $\pm$  0.00 &      294.6 $\pm$  18.9 &      -13.2 $\pm$    0.0 \\
CXOSW J104707.3+590014 & 564944 &  1.53 &      21.51 $\pm$  0.02 &      23.03 $\pm$  0.06 &      28.77 $\pm$  0.65 &      27.37 $\pm$  3.23 &     -209.0 $\pm$  0.00 &      -19.6 $\pm$    0.0 \\
CXOSW J104654.6+585804 & 562426 &  0.85 &      22.91 $\pm$  0.03 &      23.60 $\pm$  0.06 &      16.62 $\pm$  0.52 &     -40.00 $\pm$  0.00 &     -209.0 $\pm$  0.00 &      -14.7 $\pm$    0.0 \\
CXOSW J104653.3+585735 & 561949 &  1.15 &      23.49 $\pm$  0.05 &      24.78 $\pm$  0.09 &      24.81 $\pm$  0.70 &     -40.00 $\pm$  0.00 &     -209.0 $\pm$  0.00 &      -14.7 $\pm$    0.0 \\
CXOSW J104651.0+585824 & 562326 &  1.34 &      23.47 $\pm$  0.04 &      24.40 $\pm$  0.08 &      42.93 $\pm$  0.77 &      43.02 $\pm$  3.27 &     -209.0 $\pm$  0.00 &      -14.7 $\pm$    0.0 \\
CXOSW J104648.1+590103 & 563893 &  1.43 &     -25.20 $\pm$  0.00 &      25.53 $\pm$  0.14 &       8.44 $\pm$  0.44 &     -40.00 $\pm$  0.00 &     -209.0 $\pm$  0.00 &       17.0 $\pm$    4.0 \\
CXOSW J104705.3+585850 & 563861 &  4.30 &      24.28 $\pm$  0.08 &      25.21 $\pm$  0.11 &      23.18 $\pm$  0.55 &      60.23 $\pm$  3.47 &      375.6 $\pm$  19.0 &      -16.0 $\pm$    0.0 \\
CXOSW J104703.1+585622 & 561941 &  3.17 &      24.08 $\pm$  0.07 &      24.80 $\pm$  0.09 &      26.34 $\pm$  0.62 &      37.50 $\pm$  3.28 &      299.7 $\pm$  17.0 &       21.9 $\pm$    4.5 \\
CXOSW J104658.4+585914 & 563547 &  2.69 &      21.95 $\pm$  0.02 &      22.85 $\pm$  0.06 &      57.27 $\pm$  0.82 &     -40.00 $\pm$  0.00 &     -209.0 $\pm$  0.00 &       28.2 $\pm$    6.7 \\
CXOSW J104632.7+585409 & 557958 &  1.59 &     -25.20 $\pm$  0.00 &     -25.90 $\pm$  0.00 &       5.66 $\pm$  0.39 &     -40.00 $\pm$  0.00 &     -209.0 $\pm$  0.00 &       32.6 $\pm$    4.0 \\
CXOSW J104611.1+585517 & 556975 &  0.60 &      23.69 $\pm$  0.05 &      24.10 $\pm$  0.07 &      10.42 $\pm$  0.52 &     -40.00 $\pm$  0.00 &     -209.0 $\pm$  0.00 &       88.4 $\pm$    3.4 \\
CXOSW J104607.6+585602 & 557127 &  1.58 &     -25.20 $\pm$  0.00 &     -25.90 $\pm$  0.00 &      12.61 $\pm$  0.50 &     -40.00 $\pm$  0.00 &     -209.0 $\pm$  0.00 &      256.0 $\pm$    3.2 \\
CXOSW J104557.0+590000 & 558749 &  0.35 &      23.28 $\pm$  0.04 &      24.42 $\pm$  0.08 &       8.46 $\pm$  0.49 &      51.00 $\pm$  3.69 &     -209.0 $\pm$  0.00 &       47.1 $\pm$    5.5 \\
\enddata \\
\begin{minipage}{7.0in}
1: $-$ve fluxes indicate a 4$\sigma$ upper limit for the radio and
5$\sigma$ for Spitzer and optical fluxes \\
2: Distance between X-ray and SWIRE source \\
\end{minipage}
\end{deluxetable}